\documentclass[twocolumn]{aastex631}
\usepackage[utf8]{inputenc}
\usepackage{amsmath}
\usepackage{amssymb}
\usepackage{gensymb}
\usepackage{natbib}
\usepackage{xspace}
\usepackage{graphicx}
\usepackage{rotating} 
\usepackage{subfigure}
\usepackage{textcomp}
\usepackage{float}
\usepackage{hyperref}
\usepackage{savesym}
\savesymbol{tablenum}
\usepackage{siunitx}
\restoresymbol{SIX}{tablenum}

\newcommand{\e}[1]{\ensuremath{^{#1}}\xspace}

\newcommand{\NHgal}{$N_{\text{H, Gal}}$\xspace}
\newcommand{\ecut}{$E_{\text{cut}}$\xspace}

\newcommand{\pexrav}{\textsc{pexrav}\xspace}
\newcommand{\tbabs}{\textsc{TBabs}\xspace}
\newcommand{\zphabs}{\textsc{zphabs}\xspace}
\newcommand{\cutoffpl}{\textsc{cutoffpl}\xspace}
\newcommand{\zgauss}{\textsc{zgauss}\xspace}
\newcommand{\xspec}{\textsc{xspec}\xspace}

\newcommand{\xillver}{\textsc{xillver}\xspace}
\newcommand{\relxill}{\textsc{relxill}\xspace}
\newcommand{\xillvercp}{\textsc{xillverCp}\xspace}
\newcommand{\nthcomp}{\textsc{nthComp}\xspace}

\newcommand{\ka}{K$\alpha$\xspace}

\newcommand{\kt}{$kT_{e}$\xspace}
\newcommand{\gammaledd}{$\Gamma$ -- $L$/$L_{\mbox{\scriptsize Edd}}$\xspace}
\newcommand{\eddratio}{$L$/$L_{\mbox{\scriptsize Edd}}$\xspace}
\newcommand{\thetal}{$l-\theta$\xspace}

\def \mbh {$M_{\mbox{\scriptsize BH}}$\xspace}

\def \ledd {$L_{\mbox{\scriptsize Edd}}$\xspace}

\newcommand{\chisq}{$\chi^{2}$\xspace}

\newcommand{\nustar}{\textsl{NuSTAR}\xspace}
\newcommand{\bat}{\textsl{Swift}/BAT\xspace}
\newcommand{\xrt}{\textsl{Swift}/XRT\xspace}
\newcommand{\xmm}{\textsl{XMM-Newton}\xspace}
\newcommand{\chandra}{\textsl{Chandra}\xspace}

\shortauthors{Kamraj et al.}

\begin{document}

\title{X-ray Coronal Properties of \bat-Selected Seyfert 1 Active Galactic Nuclei}

\correspondingauthor{Nikita Kamraj}
\email{Contact: nkamraj@caltech.edu}

\author[0000-0002-3233-2451]{Nikita Kamraj}
\affiliation{Cahill Center for Astronomy and Astrophysics, California Institute of Technology, Pasadena, CA 91125, USA}

\author{Murray Brightman}
\affiliation{Cahill Center for Astronomy and Astrophysics, California Institute of Technology, Pasadena, CA 91125, USA}

\author{Fiona A.~Harrison}
\affiliation{Cahill Center for Astronomy and Astrophysics, California Institute of Technology, Pasadena, CA 91125, USA}

\author[0000-0003-2686-9241]{Daniel Stern}
\affiliation{Jet Propulsion Laboratory, California Institute of Technology, Pasadena, CA 91109, USA}

\author{Javier A. García}
\affiliation{Cahill Center for Astronomy and Astrophysics, California Institute of Technology, Pasadena, CA 91125, USA}
\affiliation{Dr. Karl Remeis-Observatory and Erlangen Centre for Astroparticle Physics, Sternwartstr 7, D-96049 Bamberg, Germany}

\author[0000-0003-0476-6647]{Mislav Balokovi\'c}
\affiliation{Yale Center for Astronomy \& Astrophysics, 52 Hillhouse Avenue, New Haven, CT 06511, USA}
\affiliation{Department of Physics, Yale University, PO Box 208120, New Haven, CT 06520, USA}

\author[0000-0001-5231-2645]{Claudio Ricci}
\affiliation{N\'ucleo de Astronom\'ia de la Facultad de Ingenier\'ia, Universidad Diego Portales, Av. Ej\'ercito Libertador 441, Santiago 22, Chile}
\affiliation{Kavli Institute for Astronomy and Astrophysics, Peking University, Beijing 100871, People's Republic of China}
\affiliation{George Mason University, Department of Physics \& Astronomy, MS 3F3, 4400 University Drive, Fairfax, VA 22030, USA}

\author[0000-0002-7998-9581]{Michael J. Koss}
\affiliation{Eureka Scientific, 2452 Delmer Street Suite 100, Oakland, CA 94602-3017, USA}
\affiliation{Space Science Institute, 4750 Walnut Street, Suite 205, Boulder, Colorado 80301, USA}

\author[0000-0001-8450-7463]{Julian E. Mej\'ia-Restrepo}
\affiliation{European Southern Observatory, Casilla 19001, Santiago 19, Chile}

\author[0000-0002-5037-951X]{Kyuseok Oh}
\affiliation{Korea Astronomy \& Space Science institute, 776, Daedeokdae-ro, Yuseong-gu, Daejeon 34055, Republic of Korea}
\affiliation{Department of Astronomy, Kyoto University, Kitashirakawa-Oiwake-cho, Sakyo-ku, Kyoto 606-8502, Japan}
\affiliation{JSPS Fellow}

\author[0000-0003-2284-8603]{Meredith C. Powell}
\affiliation{Kavli Institute of Particle Astrophysics and Cosmology, Stanford University, 452 Lomita Mall, Stanford, CA 94305, USA}

\author[0000-0002-0745-9792]{C. Megan Urry}
\affiliation{Yale Center for Astronomy \& Astrophysics, 52 Hillhouse Avenue, New Haven, CT 06511, USA}
\affiliation{Department of Physics, Yale University, PO Box 208120, New Haven, CT 06520, USA}

\begin{abstract}

\noindent The corona is an integral component of Active Galactic Nuclei (AGN) which produces the bulk of the X-ray emission above 1--2 keV. However, many of its physical properties and the mechanisms powering this emission remain a mystery. In particular, the temperature of the coronal plasma has been difficult to constrain for large samples of AGN, as constraints require high quality broadband X-ray spectral coverage extending above 10 keV in order to measure the high energy cutoff, which provides constraints on the combination of coronal optical depth and temperature. We present constraints on the coronal temperature for a large sample of Seyfert 1 AGN selected from the \bat survey using high quality hard X-ray data from the \nustar observatory combined with simultaneous soft X-ray data from \xrt or \xmm. When applying a physically-motivated, non-relativistic disk reflection model to the X-ray spectra, we find a mean coronal temperature \kt $=$ 84$\pm$9 keV. We find no significant correlation between the coronal cutoff energy and accretion parameters such as the Eddington ratio and black hole mass. We also do not find a statistically significant correlation between the X-ray photon index, $\Gamma$, and Eddington ratio. This calls into question the use of such relations to infer properties of supermassive black hole systems. 

\end{abstract}


\section{Introduction}\label{sec:intro}

Active Galactic Nuclei (AGN) are known to produce copious amounts of hard X-ray radiation. This continuum X-ray emission is believed to be produced in a hot cloud of plasma called the \emph{corona}, where electrons Compton up-scatter thermal optical and UV photons from the accretion disk to X-ray energies \citep[e.g.,][]{corona-haardt,merloni-2001}. While many of its physical properties are not well constrained, the corona is known to be compact, of the order of 3--10 $R_{g}$ (where $R_{g} = G$\mbh$/c^{2}$ is the gravitational radius for a black hole of mass \mbh), as determined by methods such as rapid X-ray variability \citep[e.g.,][]{mchardy-2005}, quasar microlensing \citep[e.g.,][]{chartas-lensing}, and reverberation mapping of X-ray radiation reprocessed by the accretion disk \citep[e.g.,][]{fabian-2009,demarco-2013,uttley-2014}. AGN coronae may also be compact in a radiative sense, indicating an abundance of interactions involving significant energy exchange between particles and photons within the source \citep{fabian-2015}. This radiative compactness can be characterized by the dimensionless parameter $l$ \citep{guilbert-1983}, defined as:

\begin{equation}
\label{eq:l}
 l = 4\pi \frac{m_{p}}{m_{e}} \frac{R_{g}}{R} \frac{L}{L_{E}}
\end{equation}

\noindent where $m_{p}$ and $m_{e}$ are the proton and electron mass respectively, $R_{g}$ is the gravitational radius, $R$ is the coronal radius, $L$ is the coronal luminosity, and $L_{E}$ is the Eddington luminosity.

Some of the fundamental physical properties of the corona, such as its temperature (\kt) and optical depth ($\tau$) can be probed through broadband X-ray spectroscopy. Specifically, the coronal X-ray emission can be characterized by a high-energy cutoff, \ecut, when approximating the X-ray continuum flux as a power law $\propto E^{-\Gamma}\,e^{-E/E_{\rm \tiny cut}}$, where $E$ is the photon energy and $\Gamma$ is the continuum photon index \citep[e.g.,][]{rothschild-1983}. Spectral parameters obtained from broadband X-ray fitting thus correspond to physical parameters of the corona, with the temperature related to the cutoff energy via \ecut $\sim$ 2--3 \kt, assuming a slab-like coronal geometry \citep{petrucci-2001}. Measurements of the cutoff energy have been difficult to obtain, since they require high-quality broadband X-ray spectral coverage above 10 keV. Previous studies of \ecut performed with non-focusing/collimating X-ray instruments in the hard X-ray band such as \emph{CGRO/OSSE} \citep[e.g.,][]{rothschild-1983,Zdziarski-2000}, \emph{BeppoSAX} \citep[e.g.,][]{nicastro-2000,dadina-07} and \emph{INTEGRAL} \citep[e.g.,][]{beckmann-2009,ricci-2011,malizia-2014,lubinski-2016} had limited sensitivity and could only constrain \ecut for the brightest nearby AGN.

The launch of the \nustar observatory \citep{nustar-harrison} has transformed measurements of AGN cutoff energies. Being the first focusing hard X-ray telescope in orbit with spectral coverage up to 79 keV, \nustar has enabled \ecut to be constrained for many individual unobscured and obscured AGN \citep[e.g.,][]{ballantyne-2014,brenneman-2014,mislav-2015,kara-2017,yanjun-2017}. The \bat catalog \citep[e.g.,][]{swift-mission,swift-survey,bat-105-month} provides a large sample of local, bright AGN with uniform sky coverage. Several studies have presented X-ray spectral analyses of \bat-selected AGN to investigate physical properties of the accreting supermassive black hole (SMBH). For example, \citet{ricci-2017a} performed a broadband spectral analysis of 836 \bat AGN and found a median cutoff energy for the entire sample of \ecut $=$ 200$\pm$29 keV; however, these measurements did not use \nustar data and primarily utilised lower quality hard X-ray data from non-focusing instruments such as \bat. \citet{nikita-2018} presented \ecut constraints for a sample of 46 \bat-selected Seyfert 1 AGN using $\sim$ 20 ks exposure \nustar snapshot observations performed as part of the \nustar Extragalactic Legacy Surveys program\footnote{\url{https://www.nustar.caltech.edu/page/legacy_surveys}}. More recently, \citet{balokovic-2020} presented \ecut constraints for obscured AGN selected from the \bat catalog, that also have short, 20 ks \nustar exposures. 

In addition to constraining cutoff energies, \nustar has revitalized deeper exploration of AGN coronal parameters and their possible connection with the accretion properties of SMBH systems, such as the associated Eddington ratio. While some past studies of AGN that did not utilize \nustar measurements have found tentative correlations between median values of \ecut and accretion parameters such as the Eddington ratio \citep[e.g.,][]{ricci-2018}, studies of small samples of AGN observed with \nustar have shown no evidence for such a correlation \citep{tortosa-2018}.

In this paper, we present the first systematic study of the coronal properties of a large sample of unobscured AGN observed with \nustar. We use 195 observations of Seyfert 1 AGN selected from the \bat all-sky survey that also have snapshot \nustar legacy observations or long exposure targets observed as part of individual Guest Observer programs. We include simultaneous soft X-ray data from the \xrt and \xmm instruments where available. This study provides a combination of superior quality broadband X-ray data with a large sample size, enabling robust characterization of the physical properties of the corona in the local, unobscured AGN population. 

This paper is structured as follows: in section~\ref{sec:data} we describe the sample used in this work, the X-ray observations, and data reduction procedures; in section~\ref{sec:modeling} we detail the various spectral models considered for fitting to the broadband X-ray data; in section~\ref{sec:results} we present constraints on coronal temperatures for our sample and investigate the relation between parameters derived from spectral fitting and accretion properties such as Eddington ratio and black hole mass; we summarize our findings in section~\ref{sec:summary}. We quote parameter uncertainties from spectral fitting at the 90\% confidence level.

\section{Sample, Observations and Data Reduction}\label{sec:data}

\subsection{Seyfert 1 Sample}

For this study, we selected sources by choosing AGN from the \bat 70-month X-ray catalog \citep{swift-survey}. The all-sky catalog consists of AGN that are bright in the hard X-ray band (14--195 keV). Among these sources, we select AGN that are optically classified as Seyfert 1 (Sy1), a sample that contains sub-classes ranging from Sy1--Sy1.8, following the Osterbrock classification system \citep{osterbrock-1981}. The optical spectroscopic classification is derived from the BAT AGN Spectroscopic Survey (BASS;\footnote{\url{https://www.bass-survey.com/}} \citet{koss-2017}). From this sample of unobscured AGN we then selected sources that had been observed by \nustar, both as part of the Extragalactic Legacy Survey and Guest Observer program observations. Our final sample contains 195 Sy 1 AGN with redshifts in the range $0.002 < z < 0.2$. In addition to the \nustar observations, where available we utilize simultaneous soft X-ray data in the 0.4--10 keV band taken with either the \xrt or \xmm/EPIC instruments.     

\subsection{\nustar}

We performed reduction of the raw event data from both \nustar modules, FPMA and FPMB \citep{nustar-harrison} using the \nustar Data Analysis Software (NuSTARDAS, version 2.14.1), distributed by the NASA High-Energy Astrophysics Archive Research Center (HEASARC) within the HEASOFT package, version 6.27. We calculated instrumental responses based on the \nustar calibration database (CALDB), version 20180925. We cleaned and filtered raw event data for South Atlantic Anomaly (SAA) passages using the \texttt{nupipeline} module. We then extracted source and background spectra from the calibrated and cleaned event files using the \texttt{nuproducts} module. More detailed information on these data reduction procedures can be found in the \nustar Data Analysis Software Guide \citep{nustardas}. Source spectra were extracted from circular regions with an extraction radius ranging from 30\arcsec\ to 60\arcsec\ depending on the source size and brightness. We extracted background spectra from source-free regions of the image on the same detector chip as the source, away from the outer edges of the field of view, which have systematically higher background.

\subsection{\xmm}

In addition to the \nustar hard X-ray data, where available we utilize simultaneous observations from the \xmm observatory \citep{xmm-mission} in the soft X-ray band taken with the EPIC-pn detector \citep{xmm-pn}. We have simultaneous \xmm observations for 26 observations in our sample. We performed reduction of the raw \xmm data using the \xmm Science Analysis System (SAS, \citealp{xmm-sas}, version 16.1.0), following the standard prescription outlined in the \xmm ABC online guide.\footnote{\url{https://heasarc.gsfc.nasa.gov/docs/xmm/abc/}} Calibrated, cleaned event files were created from the raw data files using the SAS command \texttt{epchain} for the EPIC-pn detector. As recommended, we only extracted single and double pixel events for EPIC-pn. Source spectra were extracted from the cleaned event files using the SAS task \texttt{xmmselect}. Background spectra were extracted from a circular aperture placed near the source on the same CCD chip. We checked for the presence of detector pileup using the SAS task \texttt{EPATPLOT}. Where significant pileup was detected in an observation we used an annular extraction region in which the core of the source point spread function is excised in order to remove pileup. We generated instrumental response files using the SAS tasks \texttt{rmfgen} and \texttt{arfgen}. 

\subsection{\xrt} 

For sources where simultaneous soft X-ray coverage with \xmm was not available, we use simultaneous data taken with the \xrt instrument \citep{swift-xrt}. We have simultaneous \xrt observations for 149 observations in our sample. We reduced the \xrt data using the ASDC XRT Online Analysis service\footnote{\url{http://www.asdc.asi.it/mmia/index.php?mission=swiftmastr}}. We performed standard filtering using the \texttt{XRTPIPELINE} script following the guidelines detailed in \citet{evans-2009}. We extracted background spectra from large annular regions around the source, taking care to avoid contamination. Instrumental ARF and RMF response files were generated using the \texttt{XRTPRODUCTS} script.   

\section{X-ray Spectral Modeling}\label{sec:modeling}

We performed all spectral modeling of the broadband X-ray data using the XSPEC fitting tool (v12.11, \citealp{xspec}). We adopt cross sections from \citet{vern} and solar abundances from \citet{wilm}. In all our model fitting we include a Galactic absorption component modeled with the \tbabs absorption code \citep{tbabs}, using Galactic column densities \NHgal taken from the HI maps of the LAB survey \citep{galactic-nh}. We also add a cross-normalization constant factor ($c_{ins}$) to all models to account for variability and calibration uncertainties across different instruments. 

\subsection{Model 1: Simple Absorbed Power Law}

\begin{figure}[t]
	\hspace{-35pt}
	\includegraphics[width=0.55\textwidth]{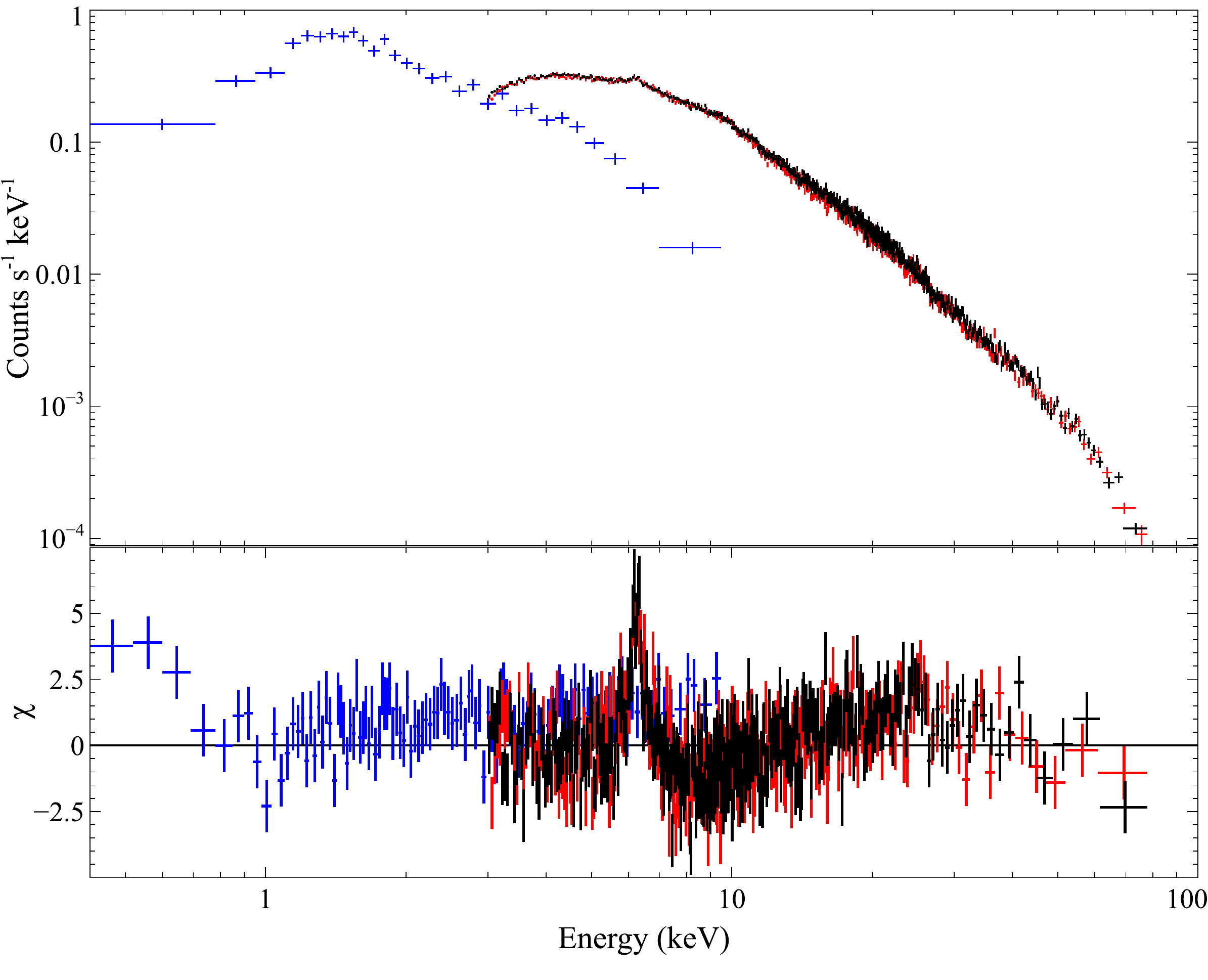}
	\caption{Broadband X-ray spectrum of IC4329A (top) alongside fit residuals for an absorbed cutoff power law model (bottom). Blue points represents \xrt data while black and red points correspond to \nustar FPMA and FPMB data respectively. Data points are re-binned for plotting clarity.}
	\vspace{5pt}
	\label{fig:specresiduals}	
\end{figure}

The first model we employ consists of a simple absorbed power law continuum with a high-energy cutoff \ecut. The power law continuum slope is characterized by the photon index $\Gamma$, with the intrinsic continuum flux proportional to $E^{-\Gamma}$exp($-E/E_{\rm cut}$). In XSPEC notation, the model is given by $c_{ins}$ $\times$ \tbabs $\times$ \zphabs $\times$ \cutoffpl, where \zphabs models photoelectric host galaxy absorption. Figure~\ref{fig:specresiduals} presents an example broadband X-ray spectrum for one of the sources in our sample, alongside residuals to the absorbed power law model fit to the data.  

\subsection{Model 2: Phenomenological Reflection (\pexrav)}

The second model we apply accounts for reflection features in the X-ray spectrum from reprocessing of the hard X-ray emission, such as the Fe \ka line at a rest-frame energy of 6.4 keV and Compton reflection hump peaking around 20--30 keV (see e.g. Figure~\ref{fig:specresiduals}). We model these features using the phenomenological \pexrav model \citep{pexrav}, which assumes reflection off a slab of semi-infinite extent and optical depth. Our model expression in XSPEC is given by $c_{ins}$ $\times$ \tbabs $\times$ \zphabs $\times$ (\cutoffpl + \zgauss + \pexrav), where \zgauss models a Gaussian Fe \ka line. We fix iron and light element abundances to solar values and fix the inclination angle of the plane of reflecting material at the default value of cos $\theta = 0.45$. We tie the photon index and normalization of the reflected power law to that of the incident power law and fix the energy of the Fe \ka line at 6.4 keV.   

\subsection{Model 3: Physical Reflection (\xillvercp) Model}

The final model we apply is an advanced reflection model that accurately models the physics of reprocessed radiation from the corona. We replace the \pexrav component with the \xillvercp model \citep{xillver}, which forms part of the \relxill family of disk reflection models \citep{relxill}. These models adopt a rich atomic database, fully calculate the angular distribution of the reflected radiation and provide various geometrical models of the illuminating coronal source. The \xillvercp model treats the coronal radiation incident on the disk as a thermally Comptonized continuum using the \nthcomp code \citep{nthcomp} which vastly improves over a simple exponential cutoff power law continuum approximation. The model also self-consistently calculates the relative strength of the reflected emission, $R$, and the ionization parameter of the accretion disk, defined as $\xi = 4\pi F_{X}/n$ where $F_{X}$ is the incident X-ray flux and $n$ is the disk density. A disk density of 10$^{15}$ cm$^{-3}$ is assumed for the \xillvercp model. Other key free parameters are the photon index $\Gamma$ and the coronal electron temperature \kt. We fix the inclination angle measured with respect to the disk normal at the default value of 30\degree. These advanced reflection models are able to provide constraints on the coronal temperature up to several hundred keV, past the limit of the \nustar detector bandpass, since the ionization state and disk structure are conditioned by the high-energy region of the spectrum, which in turn determines the observed reflection features.            

\subsection{Choice of Spectral Fitting Statistic}

In our model fitting, we consider both \chisq statistics and the Cash statistic (C-stat, \citealp{cash-1979}). While \chisq statistics have traditionally been used as the primary fit statistic for X-ray spectral fitting, its usage is appropriate only when there are sufficient photon counts in a given energy bin such that the statistical variations can be approximated by a Gaussian distribution. Rebinning of spectra to achieve a Gaussian approximation can thus wash out key features in the X-ray spectrum, such as curvature at high energies from which estimations of \ecut are made. A more appropriate fit statistic to use, particularly when dealing with low photon counts, is C-stat, which maintains Poisson counting statistics and provides unbiased parameter estimation while still resembling a \chisq statistic \citep{cstat-kaastra}. 

\begin{figure}[t]
	\hspace{-30pt}
	\includegraphics[width=0.58\textwidth]{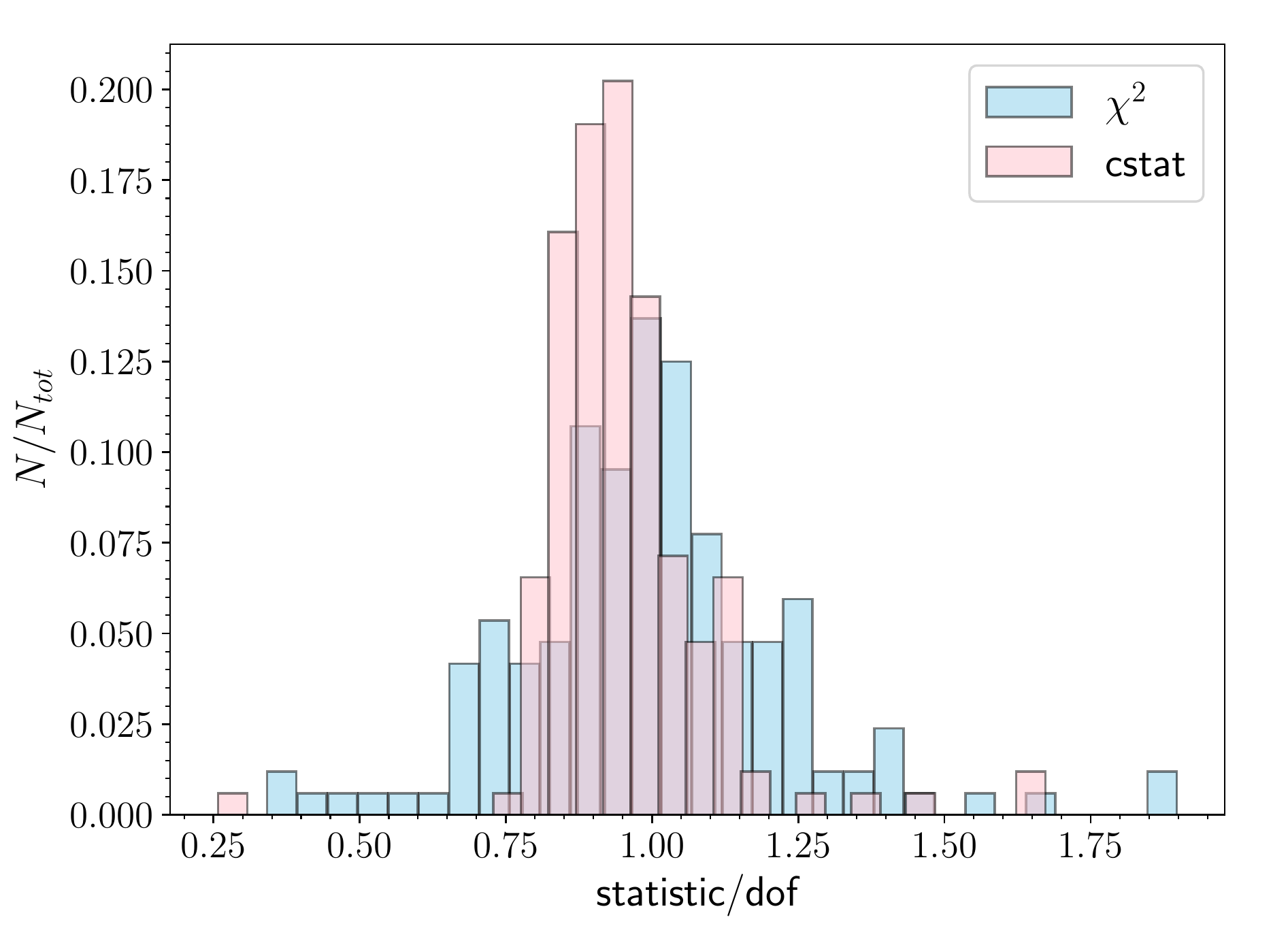}
	\caption{Distributions of the goodness-of-fit for the \chisq and Cash statistics (C-stat) for an absorbed power law model fit (Model 1) to the broadband X-ray data for our entire sample.}
	\label{fig:abspowstats}	
\end{figure}

When applying \chisq statistics, we rebinned the \nustar data to give a minimum of 20 photon counts per bin. For C-stat, we rebinned the data to have unity photon count per bin. We used the HEASOFT task \texttt{grppha} for all spectral binning. In Figure~\ref{fig:abspowstats} we show the distribution of the goodness of fit (defined as the value of the fit statistic divided by the number of degrees of freedom) for both \chisq and C-statistics, when applying the absorbed power law model fit to the data for our sample. We find that the distribution of the goodness of fit has a similar mean value for both types of statistic, indicating that C-stat provides an equally valid goodness of fit measure compared to \chisq statistics. In addition, the goodness of fit is more tightly distributed around unity for C-stat, thus providing an overall better fit, motivating its usage as the primary fit statistic for our work.   	 

\section{Results and Discussion}\label{sec:results}

\subsection{Distribution of High Energy Cutoffs}

\begin{figure}[t]
	\hspace{-8pt}
	\includegraphics[width=0.55\textwidth]{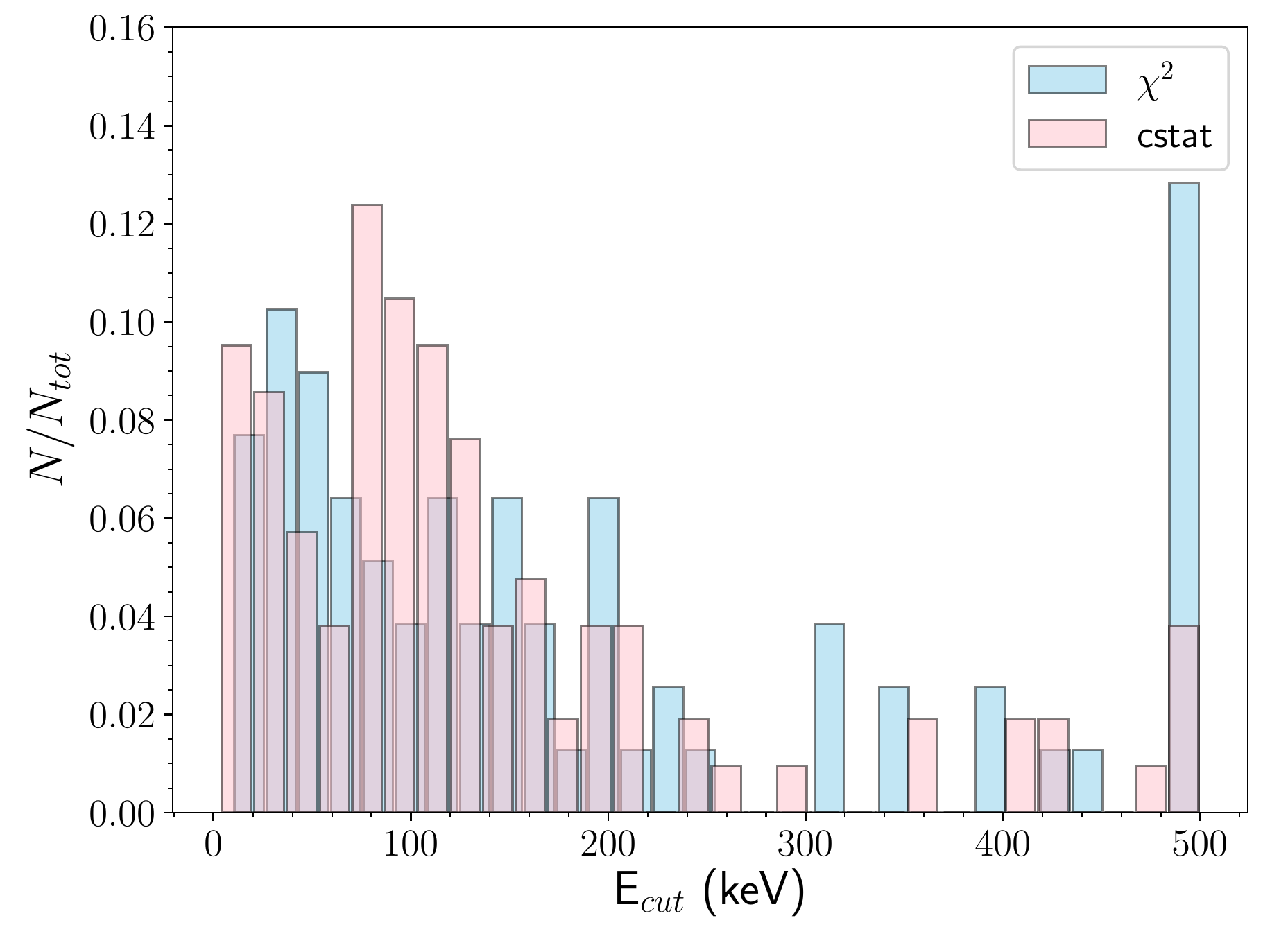}
	\vspace{5pt}
	\caption{Distributions of the high energy cutoff \ecut for the \chisq and Cash statistics (C-stat) for an absorbed power law model fit (Model 1) to the broadband X-ray data for our entire sample. We note that the model limit for the value of \ecut is 500 keV.}
	\label{fig:abspowecut}	
\end{figure} 

We examined the distributions of coronal cutoff energies determined from broadband X-ray modeling (\ecut or \kt depending on type of spectral model) for our full sample, considering both best-fit values and lower bounds. We first compared best-fit results for the \chisq and C-stat distributions and overall we find improved constraints when using C-stat. Figure~\ref{fig:abspowecut} shows the distributions of best-fit \ecut values for the absorbed power law model fit to the data for both \chisq and C-stat. We observe that with the \chisq statistic, a large number of sources have high energy cutoffs pegged at the upper limit of the model, leading to mean and median \ecut values that are skewed to higher energies. With C-stat, significantly fewer sources have best-fit coronal cutoff values at the upper limit of the model, producing average cutoffs that are lower and more closely resembling the true mean of the distribution, particularly since curvature in the spectrum at high energies associated with the coronal cutoff is not washed out, which can often be the case with the relatively wide energy bins necessary for \chisq fitting. 

We find full constraints on \ecut (both lower and upper limits) for 60 observations in our sample using C-stat, compared to 44 full \ecut constraints when performing spectral fitting with \chisq for the absorbed power law model. For the \pexrav model, we obtain full \ecut constraints for 91 observations with C-stat and 70 observations with \chisq statistics. For the final \xillvercp model, full constraints on \kt are obtained for 79 observations with C-stat and 43 observations with \chisq. In all cases, the number of full constraints on \ecut or \kt is greater when applying the C-statistic compared to the \chisq statistic. Furthermore, we note that while some individual AGN may have very high cutoff energies, e.g. Cen A \citep{felix-cenA-2016}, generally average high energy cutoffs of the AGN population cannot exceed several hundred keV as this would otherwise overproduce the cosmic X-ray background above 100 keV \citep{x-ray-background, ananna-2019}. 

\begin{figure*}[t]
	\gridline{\fig{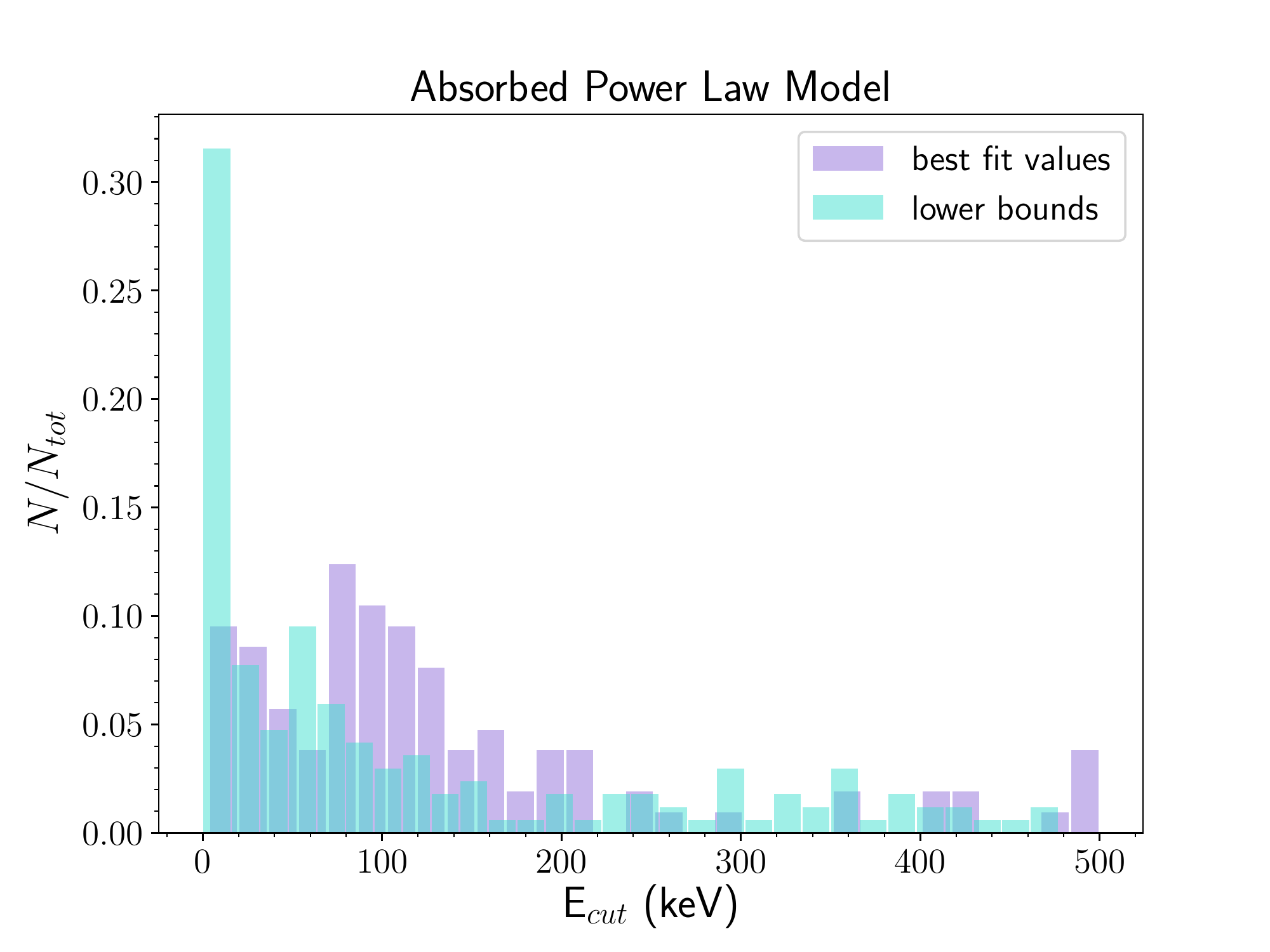}{0.52\textwidth}{(a)}
	\fig{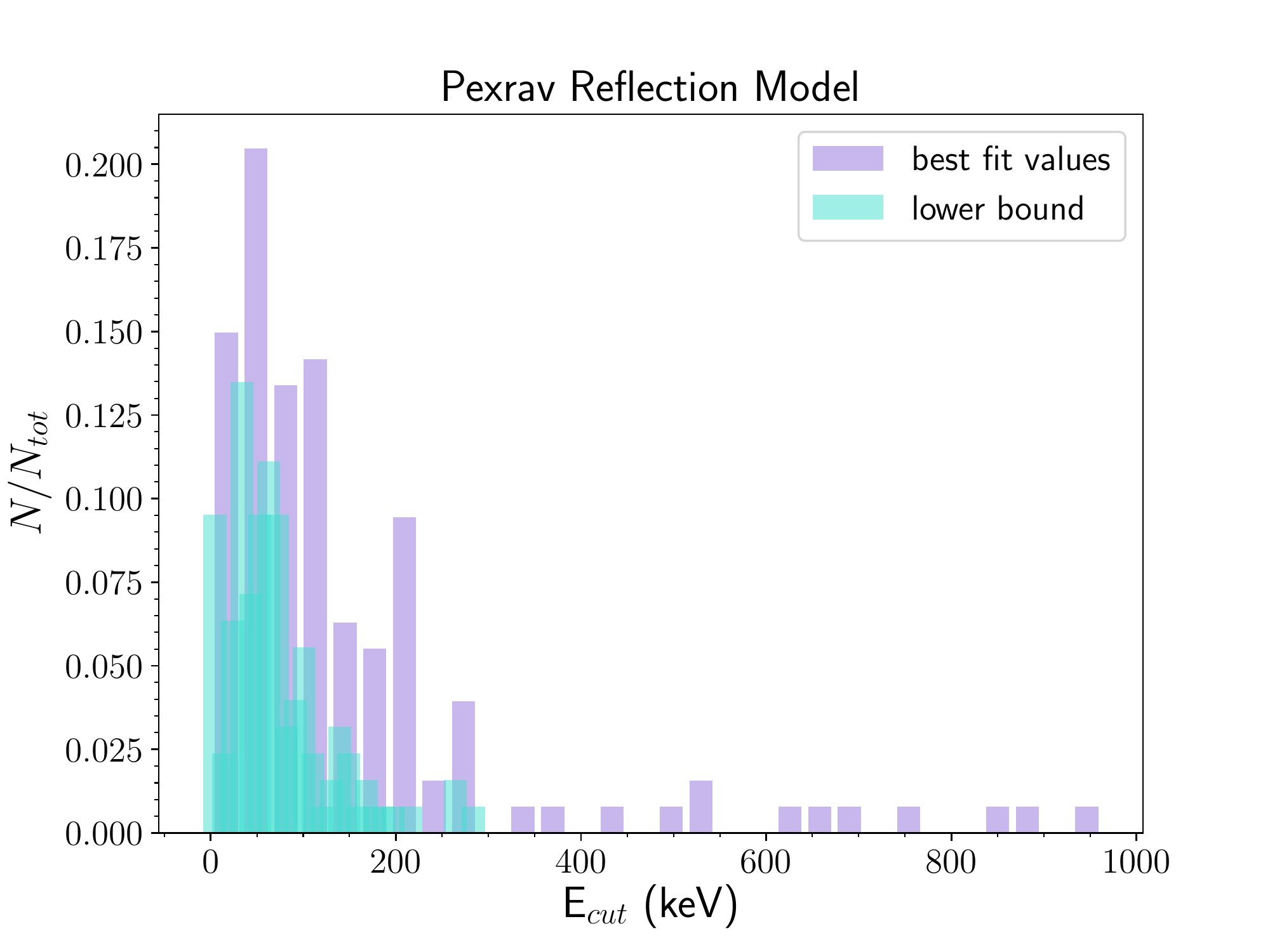}{0.52\textwidth}{(b)}
	}
	\vspace{-10pt}
	\gridline{\fig{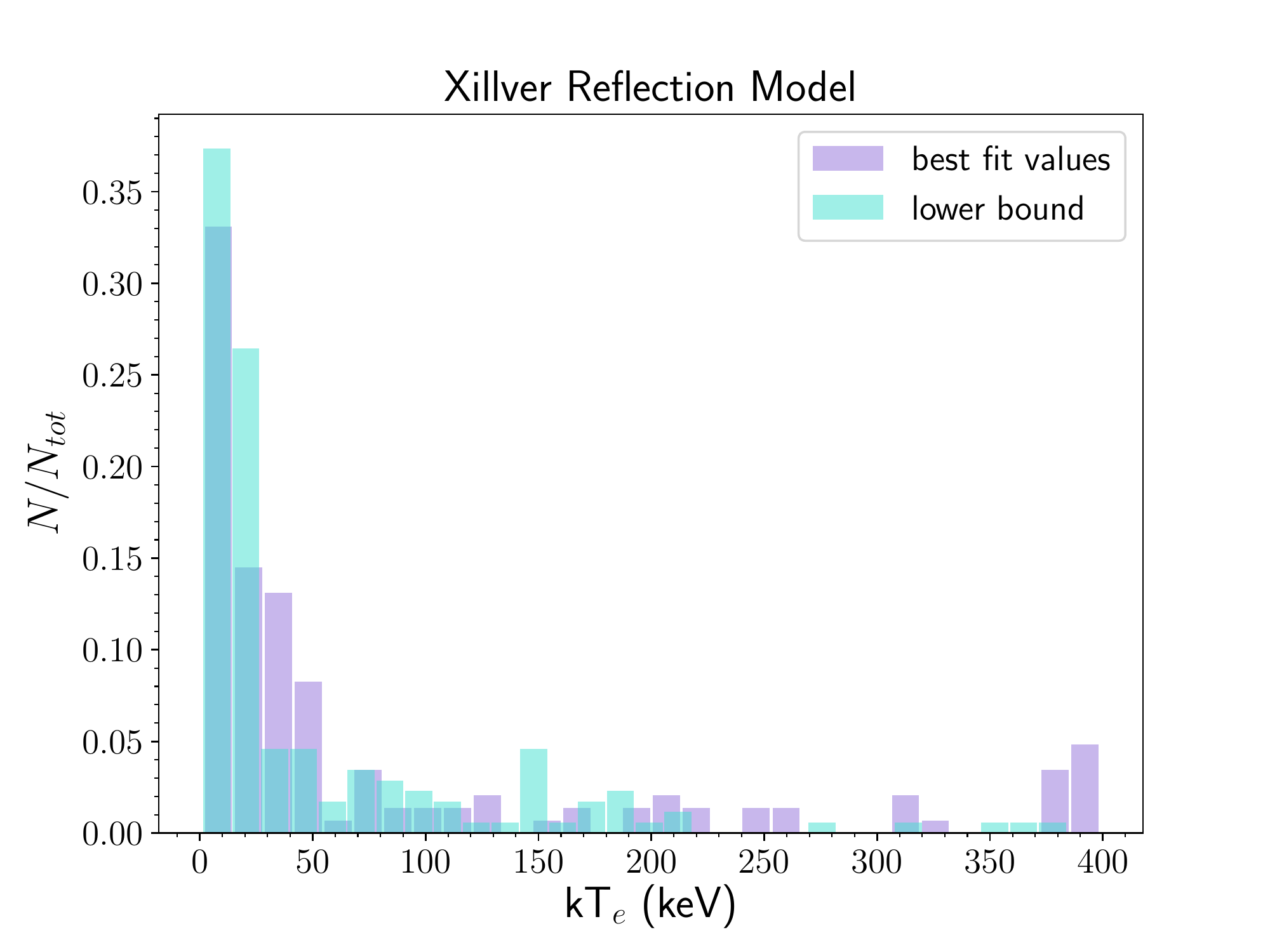}{0.52\textwidth}{(c)}
	}
	\caption{Distributions of the best-fit values and lower bounds of the high energy cutoff, \ecut, and coronal temperature, \kt, for the different spectral models applied to the broadband X-ray data for our entire sample, using C-stat. Mean best-fit values of the coronal cutoff for the different spectral models, as determined from the purple histograms, are as follows: (a) \ecut $=$ 137$\pm$12 keV (b) \ecut $=$ 156$\pm$13 keV (c) \kt $=$ 84$\pm$9 keV.}
	\vspace{5pt}
	\label{fig:specmodelsecutdist}	
\end{figure*}
 
In Figure~\ref{fig:specmodelsecutdist} we present distributions of coronal cutoff energies using C-stat for the three spectral models considered in this work. Overall, we find good agreement between the different spectral models for the mean value of \ecut. We determine the mean value of \ecut or \kt from the best-fit distributions presented in Figure~\ref{fig:specmodelsecutdist} (purple). We do not include best-fit values of \ecut or \kt that are pegged at the upper limit of the spectral model (500 keV, 1 MeV and 400 keV for models 1, 2 and 3 respectively) in our determination of mean values, as such values typically result from \xspec being unable to determine error bounds from the data. For the \xillvercp model, assuming a conversion factor for \ecut $\sim$ 2--3 \kt \citep{petrucci-2001}, the mean value of \kt$=$ 84$\pm$9 keV is consistent with the mean value of \ecut derived for spectral models 1 and 2, which were found to be 137$\pm$12 keV and 156$\pm$13 keV respectively. Table~\ref{tab:specfitresults} presents example spectral fit constraints and source properties for some select objects from our sample.

We generally find good agreement when comparing our \ecut or \kt results to similar measurements for unobscured AGN reported in the literature. For example, \citet{ricci-2017a} reported a median value of \ecut $=$ 210$\pm$36 keV for unobscured AGN from the \bat 70-month catalog, determined from broadband spectral fitting of \bat and \emph{Suzaku} data. In recent work by \citet{molina-2019} which studied \nustar spectra for a small sample of 18 broad-lined AGN selected with \emph{INTEGRAL}, a mean value of \ecut $=$ 111$\pm$45 keV was obtained for their sample, which is marginally consistent with our measurements given the large uncertainty.

\begin{table*}
	\centering	
	\caption{Redshifts, black hole masses and best-fit spectral parameters from fitting broadband X-ray data for select sources from our sample of \bat-selected Sy1s}
	\renewcommand{\arraystretch}{1.0}
	\label{tab:specfitresults}
	\begin{tabular*}{\textwidth}{@{\extracolsep{\fill} }l c c c c c c c c}
		
		\noalign{\smallskip} \hline \hline \noalign{\smallskip}
		
		Source & Redshift & log($M_{\rm BH}/M_{\bigodot}$)$^{A}$ & $\Gamma$ & \ecut & \kt & C-stat/dof & Model$^{B}$ \\
		& & & & (keV) & (keV) & & \\ \hline
		
		Fairall 1203 & 0.058 & 8.20 & 1.24$^{+0.15}_{-0.17}$ & 42$^{+51}_{-16}$ & - & 1052/1192 & 1 \\
        & & & 1.45$^{+0.23}_{-0.19}$ & $\geq$ 45 & - & 186/247 & 2\\ 
		& & & 1.56$^{+0.05}_{-0.04}$ & - & 15$^{+39}_{-4}$ & 1043.5/1189 & 3\\
		
		IC 4329A & 0.016 & 7.81 & 1.69$^{+0.004}_{-0.01}$ & $\geq$ 394 & - & 4955/4179 & 1 \\
        & & & 1.72$\pm$0.01 & 170$^{+23}_{-18}$ & - & 2698/2706 & 2\\ 
		& & & 1.77$^{+0.01}_{-0.004}$ & - & 82$^{+16}_{-7}$ & 4275.7/4210 & 3\\
		
		MCG-3-4-72 & 0.046 & 7.09 & 1.73$\pm$0.05 & 117$^{+223}_{-48}$ & - & 1554/1642 & 1 \\
		& & & 1.95$\pm$0.06 & 123$^{+596}_{-59}$ & - & 748.5/749 & 2\\ 
		& & & 1.73$^{+0.04}_{-0.05}$ & - & 15$^{+23}_{-4}$ & 1537/1654 & 3\\
        
		Mrk 1148 & 0.064 & 8.01 & 1.73$^{+0.05}_{-0.04}$ & 81$^{+59}_{-18}$ & - & 2138/2262 & 1 \\
		& & & 1.89$\pm$0.04 & $\geq$ 85 & - & 1477/1633 & 2\\
        & & & 1.78$^{+0.04}_{-0.03}$ & - & $\geq$ 18 & 2167.5/2290 & 3\\ 
		
		RBS 542 & 0.104 & 7.89 & 1.61$\pm$0.02 & 74$^{+14}_{-10}$ & - & 3389.8/3251 & 1 \\
        & & & 1.64$\pm$0.03 & 49$^{+7}_{-5}$ & - & 1658/1570 & 2\\ 
		& & & 1.70$^{+0.03}_{-0.02}$ & - & $\geq$ 16 & 3253/3257 & 3\\

\\ \hline	
		
	\end{tabular*}
\raggedright{\tablecomments{Information presented in this table is available for our full sample in machine-readable form. \\ $^{A}$ Reference: Mejía-Restrepo et al., submitted (BASS XXII DR2: Broad-line based black hole mass estimates and biases from obscuration). \\ $^{B}$ Applied XSPEC models: (1) $c_{ins}$ $\times$ \tbabs $\times$ \zphabs $\times$ \cutoffpl \\ (2) $c_{ins}$ $\times$ \tbabs $\times$ \zphabs $\times$ (\cutoffpl + \zgauss + \pexrav) \\ (3) $c_{ins}$ $\times$ \tbabs $\times$ \zphabs $\times$ (\xillvercp + \zgauss)}}	
\end{table*}

\subsection{Relation between the High Energy Cutoff and Accretion Parameters}

\begin{figure}[t]
    \includegraphics[width=1.0\columnwidth]{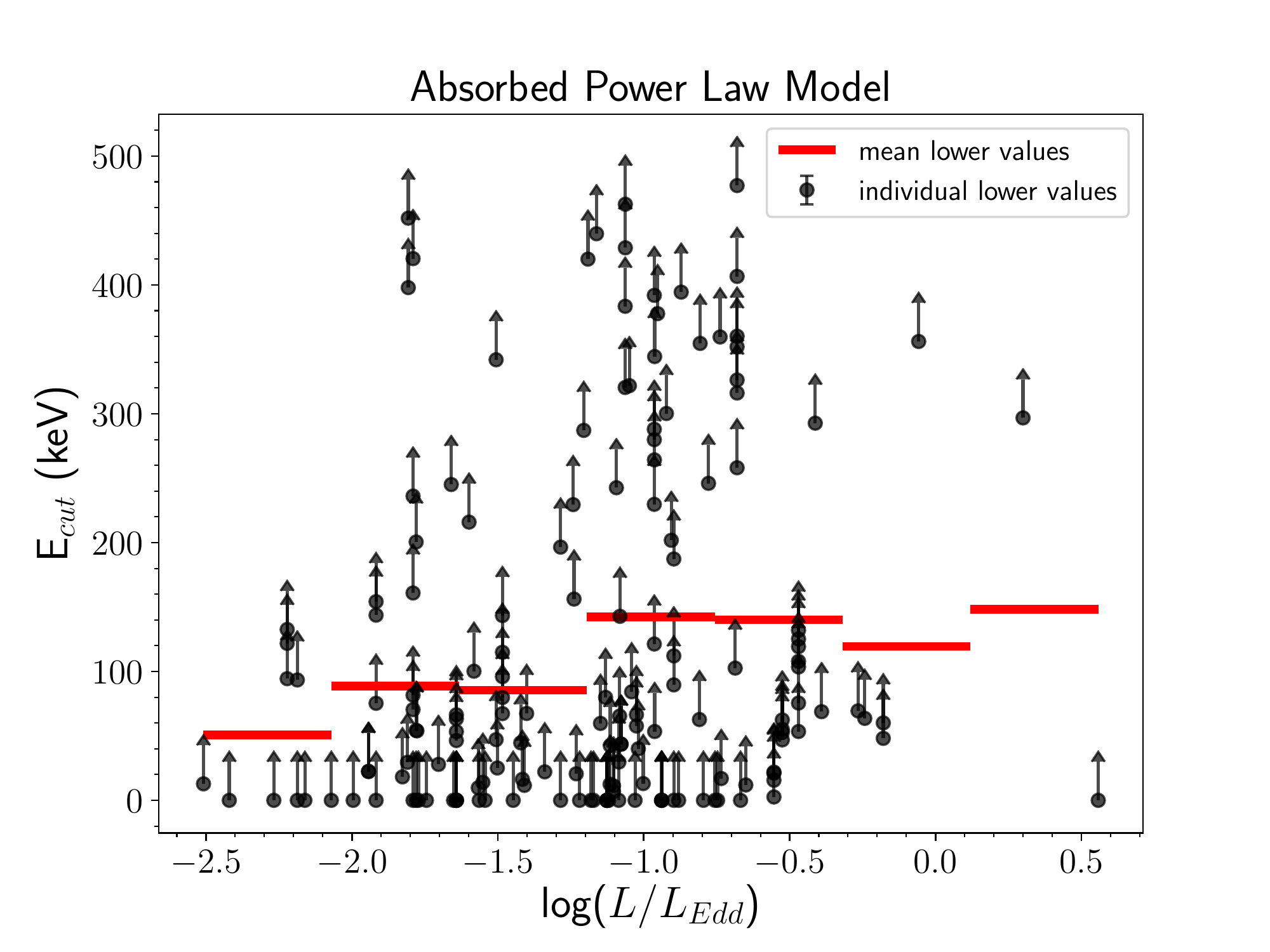}
    \includegraphics[width=1.0\columnwidth]{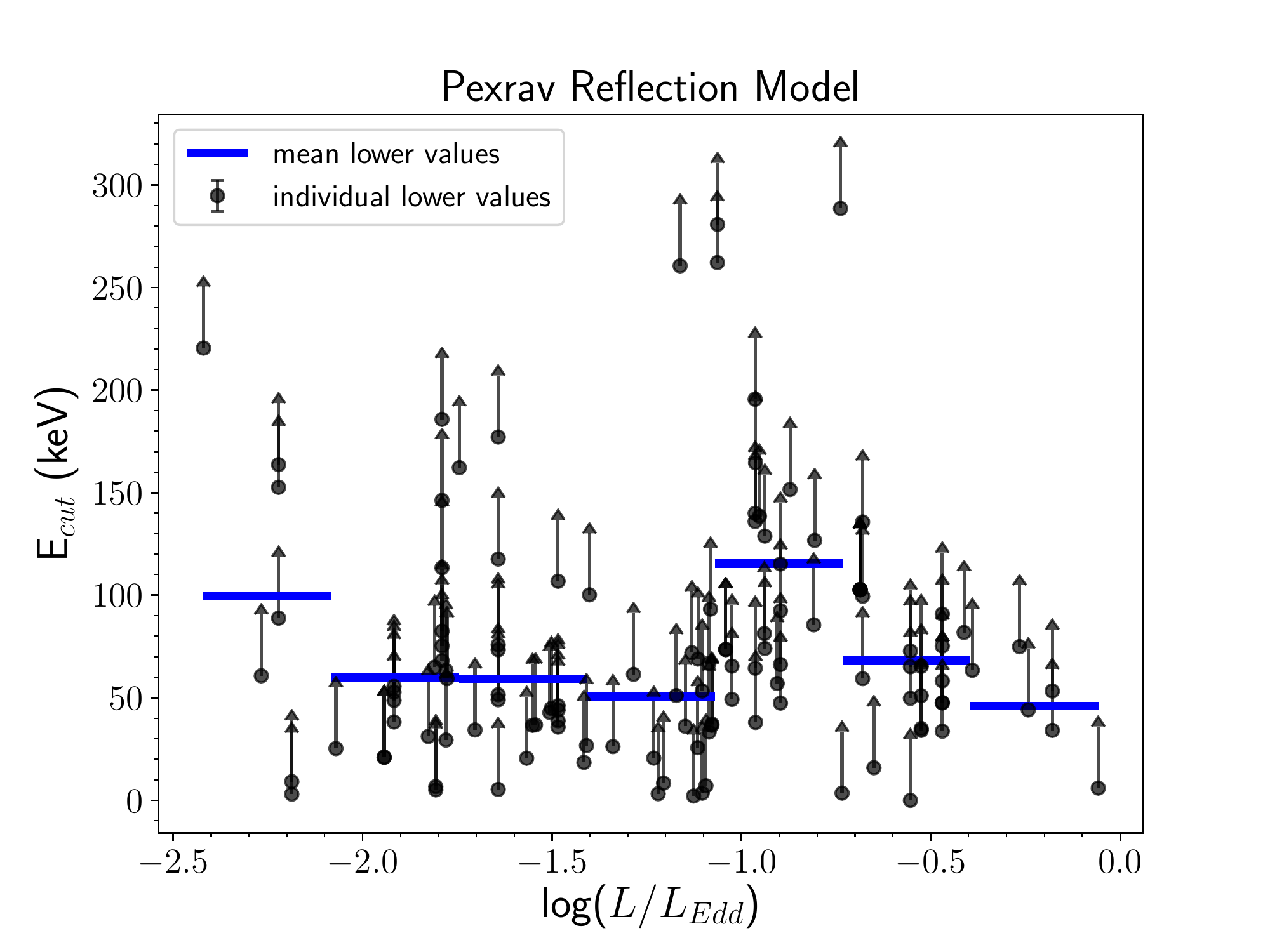}
    \includegraphics[width=1.0\columnwidth]{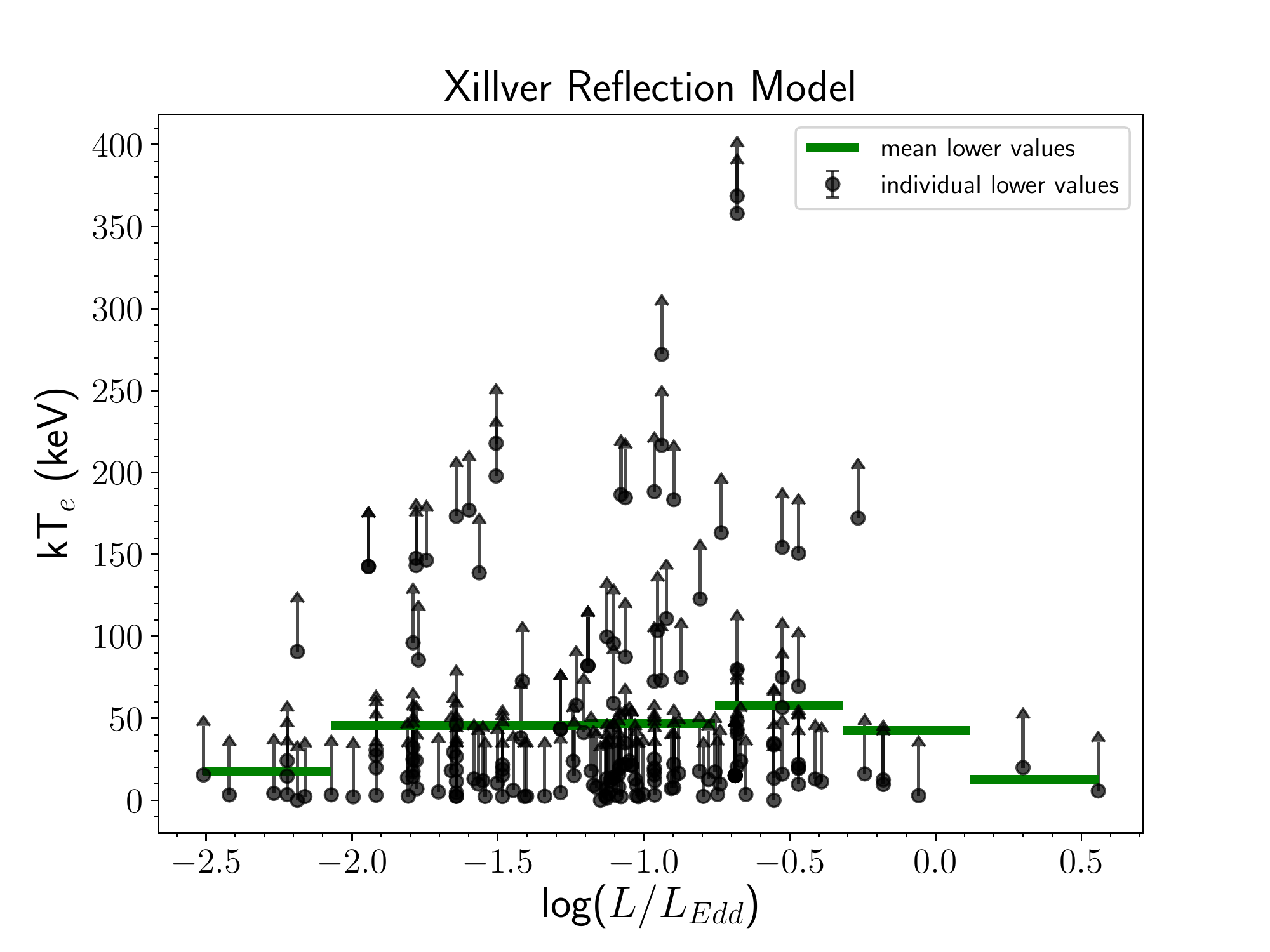}
	\caption{Lower bounds of \ecut and \kt versus the Eddington ratio \eddratio for different X-ray spectral models applied to our entire sample, using C-stat. Solid horizontal lines represent mean values of \ecut or \kt for different intervals of \eddratio.}
	\vspace{5pt}
	\label{fig:ecutvsledd}	
\end{figure}

\begin{figure}[t]
    \includegraphics[width=1.0\columnwidth]{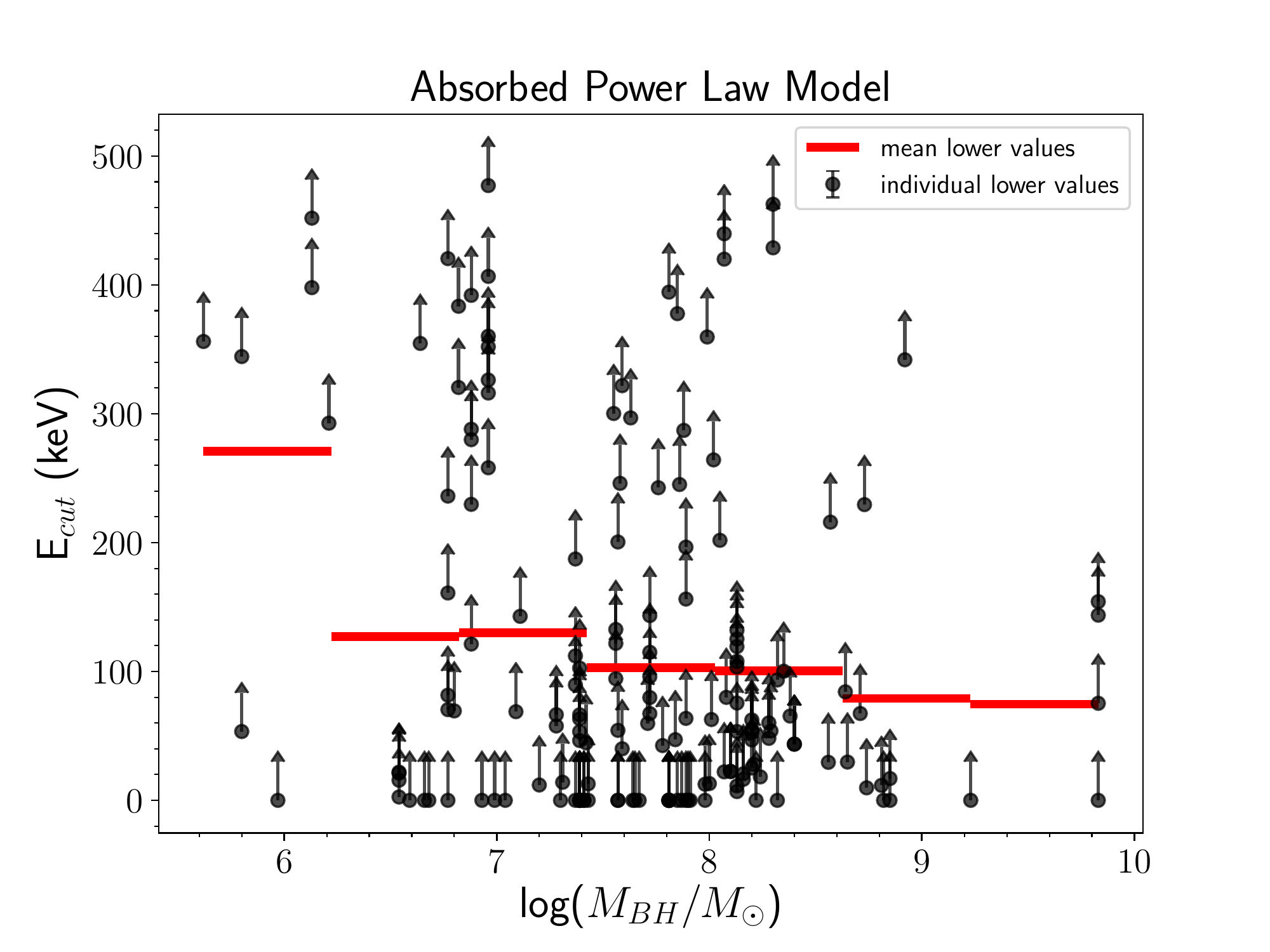}
    \includegraphics[width=1.0\columnwidth]{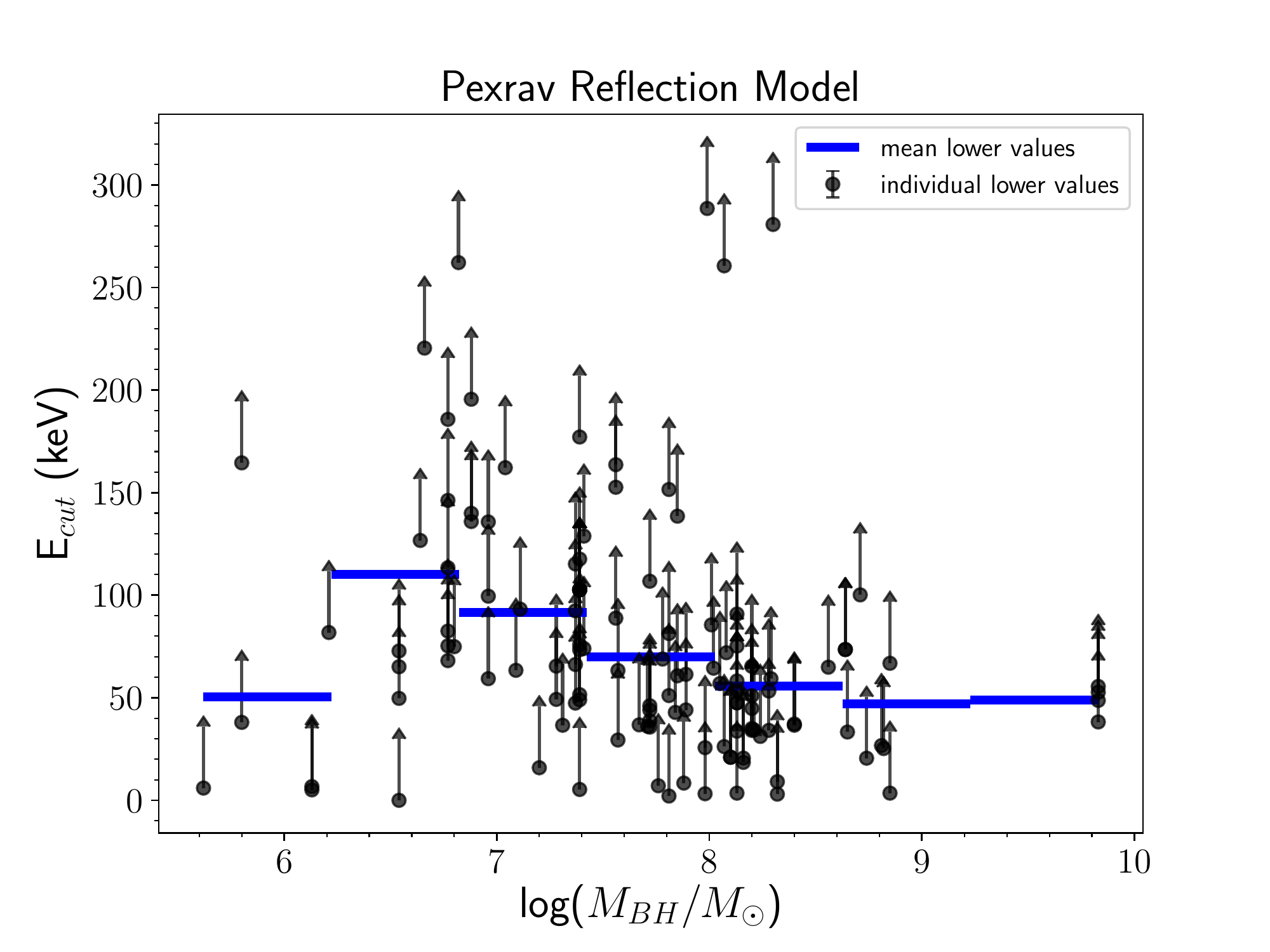}
    \includegraphics[width=1.0\columnwidth]{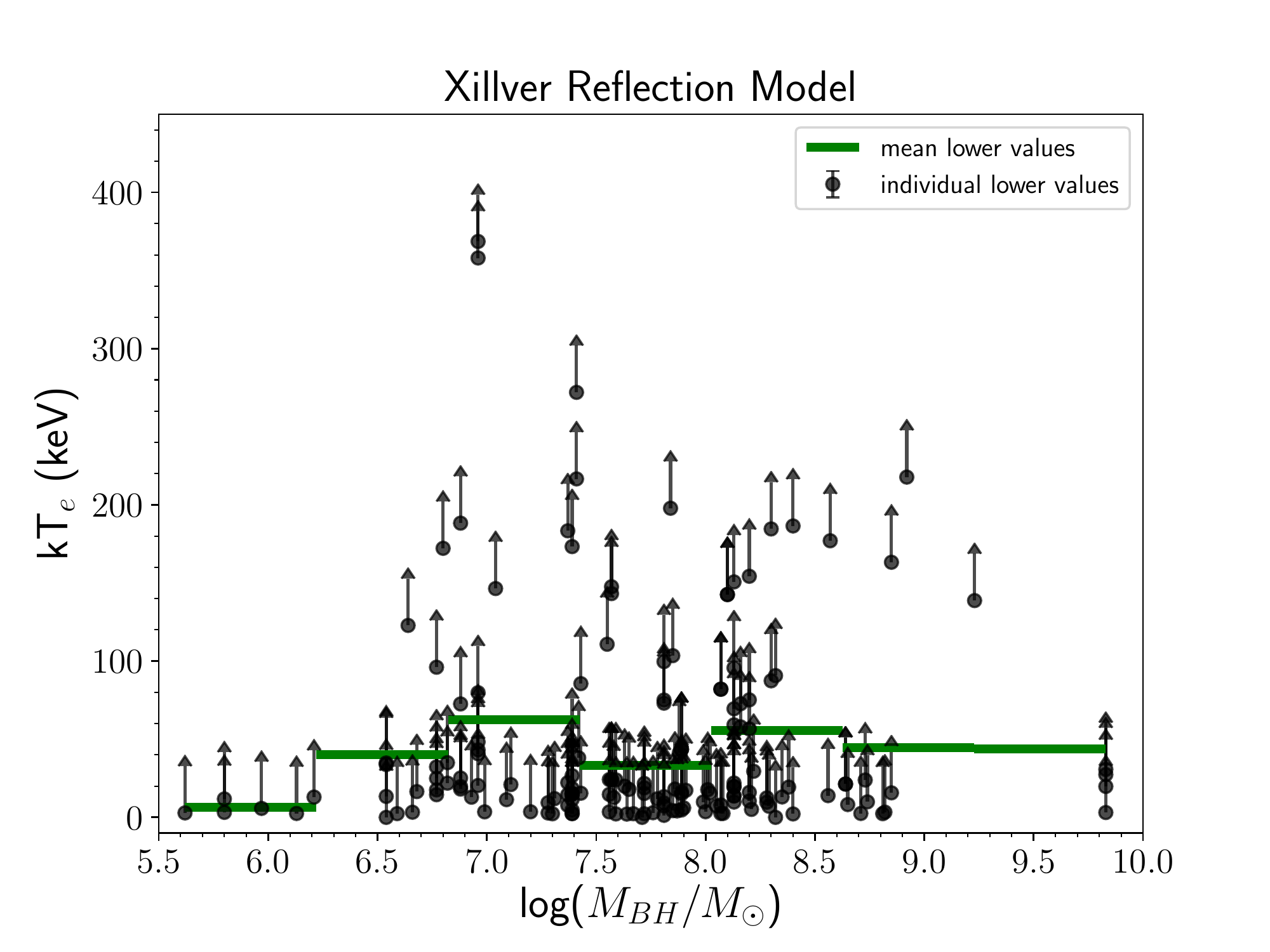}
	\caption{Lower bounds of \ecut and \kt versus SMBH mass \mbh for different X-ray spectral models applied to our entire sample, using C-stat. Solid horizontal lines represent mean values of \ecut or \kt for different intervals of \mbh.}
	\vspace{5pt}
	\label{fig:ecutvsmbh}	
\end{figure}

We next explore possible relations between the cutoff energy/temperature and fundamental accretion properties of the SMBH, such as Eddington ratio (\eddratio) and mass of the SMBH (\mbh). We adopt black hole mass estimates from the second data release of optical measurements from the BASS survey (DR2: Broad-Line Based Black Hole Mass Estimates and Biases from Obscuration, Mejía-Restrepo et al., submitted). In addition to broad-line measurements from optical spectra, black hole mass estimates from the BASS survey have been compiled using several other techniques, such as stellar velocity dispersions, direct methods such as maser emission and reverberation mapping, and the $M_{\rm BH}-\sigma_{\ast}$ relation of \citet{m-sig-relation}. The Eddington luminosity $L_{\rm Edd}$ for pure H composition is given by:

\begin{equation}
\centering
L_{\rm Edd} = \frac{4\pi GM_{BH}m_{p}c}{\sigma_{T}}
\label{eq:ledd}
\end{equation}

\noindent where $G$ is the gravitational constant, $c$ is the speed of light, and $\sigma_{T}$ is the Thompson scattering cross section. We calculate the Eddington ratio \eddratio using the bolometric AGN luminosity, determined from the intrinsic X-ray luminosity in the 2--10 keV band, applying a bolometric correction factor of 20 to the 2--10 keV luminosity \citep{bolometric-correction}. 

Figure~\ref{fig:ecutvsledd} presents the lower bound of \ecut or \kt versus Eddington ratio for the three spectral models considered. Since our sample contains a large number of sources with either partial constraints or lower limits on the coronal cutoff energy, for consistency we examine the trend in the lower bounds of \ecut or \kt across the entire sample. While our sample includes some strong constraints for bright AGN with long \nustar exposures, as noted in \citet{balokovic-2020}, for large sample statistics, such lower bound constraints are more informative than analysis of a small number of tightly constrained limits. Recently, \citet{akylas-2021} studied the coronal cutoffs in \bat-selected Sy 1 AGN using \nustar data, incorporating techniques similar to \citet{balokovic-2020} that allow them to take the numerous lower limits on \ecut into account. Best-fit values of \ecut or \kt obtained from spectral fitting with \xspec can fluctuate between the upper and lower bounds with refitting, thus the lower bound presents a more robust constraint. We also note lower bounds on the cutoff energy that lie well outside the \nustar bandpass should be considered with due caution, due to uncertainties in spectral modeling combined with limitations in data quality for some sources. For example, \citet{matt-2015} found \ecut $=$ 720$^{+130}_{-190}$ keV when fitting the \nustar spectrum of NGC 5506 with a relativistic reflection model. However, \citet{balokovic-2020} found \ecut $=$ 110$\pm$10 keV for NGC 5506 when fitting the same \nustar data with a slightly different spectral model. Despite the improved sensitivity of broadband spectra taken with \nustar and usage of more physically accurate reflection models, the extrapolation of Comptonization spectra to energies above the bandpass of \nustar presents large uncertainties and are overall difficult to predict above 100 keV \citep[e.g.,][]{zdziarski-2019,zdziarski-2021}.

We do not find a significant correlation between the lower bound of the coronal cutoff and Eddington ratio for the full sample, with p-values exceeding 1 \% when applying a Spearman rank correlation test to both individual points and binned data. We also investigated the dependence of the coronal cutoff on the mass of the AGN SMBH (\mbh). Figure~\ref{fig:ecutvsmbh} shows the results for different spectral models. While we find a declining trend in the mean lower bound of \ecut with \mbh for the absorbed power law model, with a p-value of 0.004\%, we do not find any statistically significant correlation for the two reflection models (p-value $>$ 10\%), which more accurately characterize the AGN emission compared to the simple power law fit.  

Comparing our results with similar studies of coronal parameters by \citet{ricci-2018} using pre-\nustar broadband X-ray data, we find some differences. \citet{ricci-2018} find a negative trend in the median value of \ecut with Eddington ratio for their sample of 211 unobscured AGN, including lower limits. However no significant correlation is found between \ecut and Eddington ratio when lower limits are ignored in their sample. While they observe a possible trend between \ecut and \mbh, such a correlation disappears when dividing the sources into bins of different Eddington ratio. \citet{tortosa-2018} studied coronal parameters in a sample of 19 \bat-selected Seyfert 1 galaxies observed with \nustar, and found no correlation between the high energy cutoff and SMBH mass or Eddington ratio. In another study, \citet{molina-2019} examined X-ray spectra of 18 broad-lined AGN selected with \emph{INTEGRAL} and observed with \nustar. They found no correlation between the high energy coronal cutoff and Eddington ratio. In a more recent study  by \citet{hinkle-2021}, which examined coronal parameters in a sample of 33 \bat-selected Sy1 and Sy2 AGN observed with \nustar and \xmm, no strong correlation was found between cutoff energy and SMBH mass or Eddington ratio. Therefore, we demonstrate that when fitting high quality broadband X-ray spectra obtained with \nustar for large samples of AGN, we do not find a strong correlation between the coronal cutoff and AGN accretion parameters such as Eddington ratio and SMBH mass.

\subsection{The {\gammaledd} Relation}

Next, we investigate the relationship between the X-ray photon index and Eddington ratio, hereafter referred to as the \gammaledd relation. Numerous studies report a positive linear correlation between these two parameters, of the form 

\begin{equation}
\centering
\Gamma = \Psi log(\frac{L}{L_{Edd}}) + \omega.
\label{eq:gamma-ledd}
\vspace{8pt}
\end{equation}

Early studies of individual sources and small samples of AGN hinted at a correlation between $\Gamma$ and \ledd \citep[e.g.,][]{pounds-1995,brandt-1997}, though such studies probed a very limited range of AGN luminosities. Studies covering a wider range of luminosities and redshifts such as \citet{shemmer-2006, shemmer-2008} and \citet{brightman-2013} identified a statistically significant correlation between $\Gamma$ and \ledd, with a slope $\Psi$ $\sim$ 0.3. However, \citet{sobolewska-2009} performed detailed spectral analysis of 10 \emph{RXTE}-observed AGN and found $\Psi$ to vary from object to object, with a flatter average slope $\Psi$ $=$ 0.08. \citet{yang-2015} constructed a large sample of AGN and black hole binaries covering a wide luminosity range, and found $\Gamma$ to be constant at very low luminosities, but varied from being positively and negatively correlated over certain luminosity ranges. Recent studies by \citet{benny-bh-paper} which examined the \gammaledd relation for 228 \bat AGN, have also reported flatter slopes ($\Psi$ $\sim$ 0.15), with overall large scatter in the relation. Furthermore, the authors found no evidence for a \gammaledd correlation for subsets of AGN with reliable, direct BH mass estimates. \citet{ricci-2013} also found a similarly flat slope ($\Psi$ $=$ 0.12) for their sample of 36 \chandra-observed AGN. In \citet{benny-bh-paper}, the authors only recover the steeper slope consistent with earlier studies ($\Psi$ $\sim$ 0.3) when applying a simple power law model fit to the subset of broad-lined AGN in their sample. These results demonstrate that the \gammaledd relation may not be robust or universal, with the strength of the correlation varying with choice of sample, luminosity ranges of the sample, energy range of X-ray data used in analysis, and type of X-ray spectral model used in determining $\Gamma$. 

\begin{figure}[t]
	\hspace{-30pt}
	\includegraphics[width=0.55\textwidth]{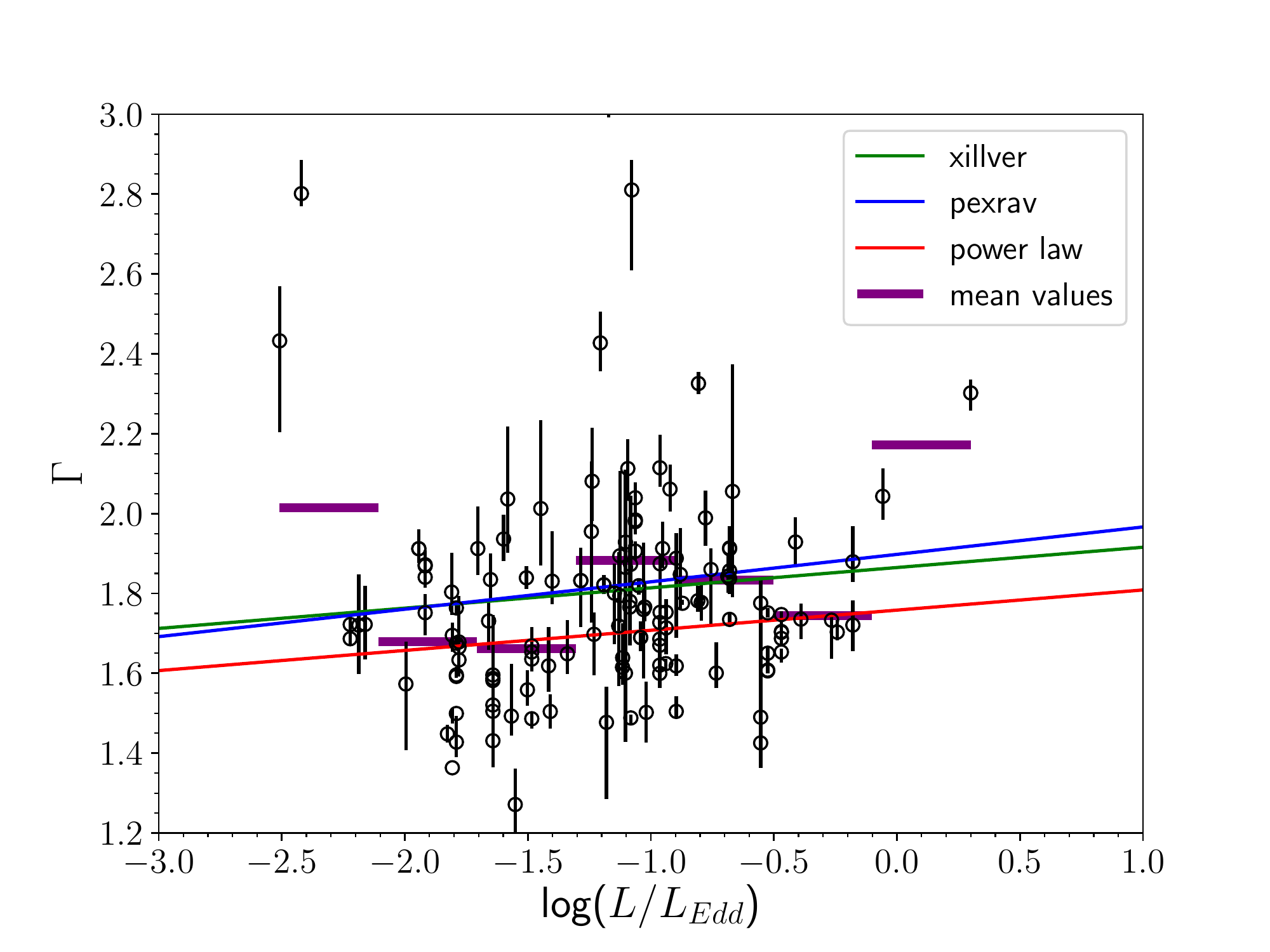}
	\vspace{5pt}
	\caption{The \gammaledd relation for our full sample. Solid lines show linear fits to the data for the three different X-ray spectral models applied in this work. Marked in purple are mean values of $\Gamma$ for different bins of Eddington ratio for the \xillver reflection model.}
	\label{fig:gammaledd}	
\end{figure}

In our work, we examined the \gammaledd relation for our full sample for all three X-ray spectral models used in our analysis. We also investigated whether there was a dependence of the slope of the relation on the type of X-ray spectral model fitted to the broadband data. We present our results for the \gammaledd relation in Figure~\ref{fig:gammaledd}, showing $\Gamma$ values for the \xillver reflection model, which most accurately characterizes the AGN X-ray emission. Overall, we find considerable scatter and no strong trend between $\Gamma$ and \ledd for all spectral models that we applied. Applying a formal Spearman rank test confirmed the absence of a statistically significant correlation, with correlation coefficients less that 0.25 and p-values exceeding 0.2\%. We find no strong correlation when dividing the data into bins of Eddington ratio (purple lines in Figure~\ref{fig:gammaledd}). When applying a simple linear regression fit to the data for each X-ray spectral model, we obtain very flat slopes: $\Psi$ $\sim$ 0.03 -- 0.06. Comparing to the literature, our slopes for the \gammaledd relation are much flatter than previously reported results, as we find little evidence for a correlation between $\Gamma$ and \ledd. We conclude that when analyzing a large, unbiased sample of unobscured AGN with high sensitivity broadband X-ray spectral data, we do not find robust evidence for a \gammaledd correlation and caution on the usage of such a relation to derive estimates of \ledd or \mbh.        

\subsection{Location of Sources in the Compactness--Temperature Plane}

\begin{figure*}[t!]
	\hspace{60pt}
    \includegraphics[width=0.7\textwidth]{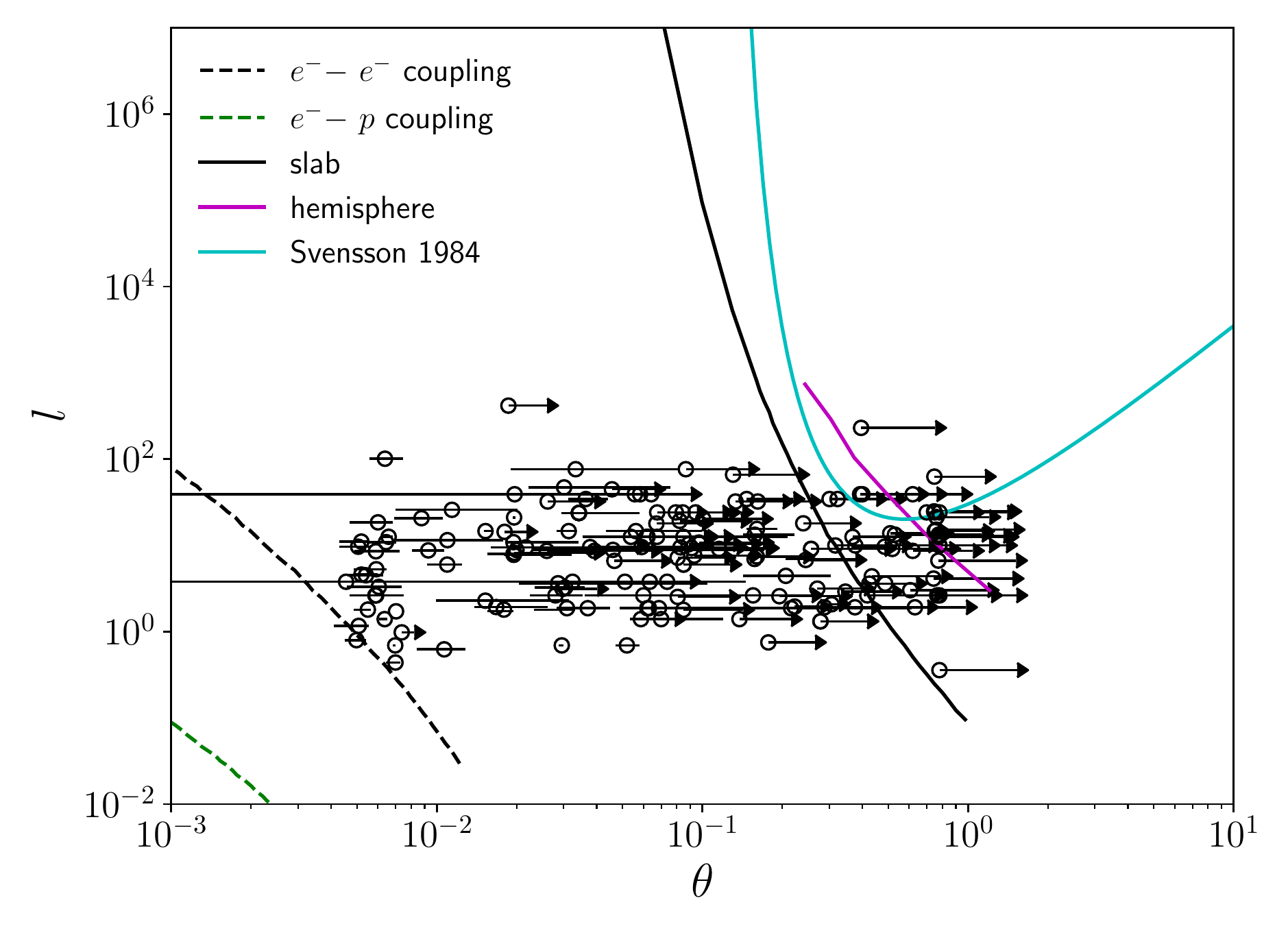}

    \hspace{60pt}
    \includegraphics[width=0.7\textwidth]{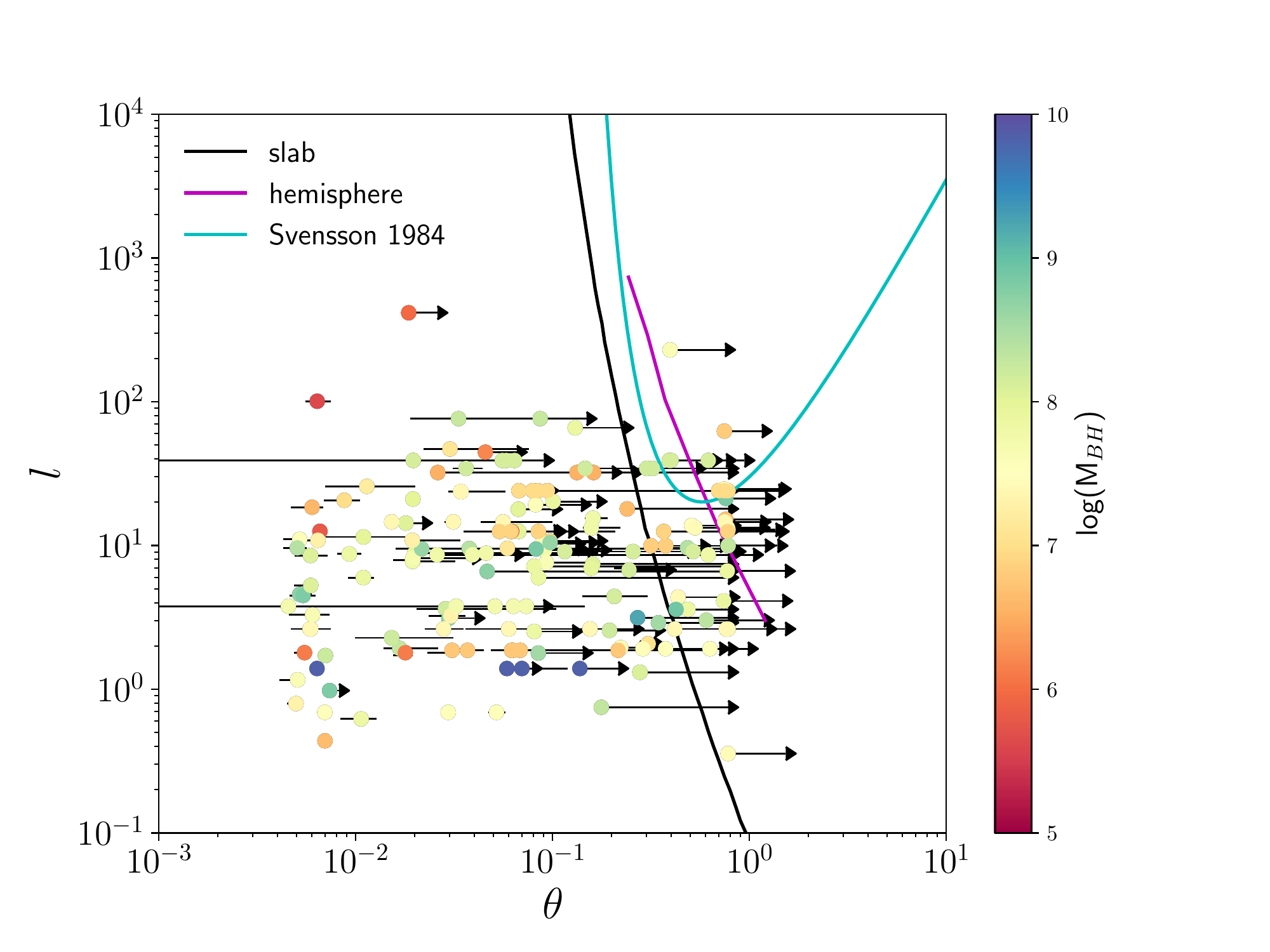}
	
	\caption{Compactness-Temperature (\thetal) diagrams for our full sample, with $\theta$ determined using \kt values from the \xillver model fit to data. Solid lines correspond to pair production lines for different coronal geometries. The dashed lines in the top panel correspond to boundaries where regions are dominated by electron-electron and electron-proton coupling processes. Data points in the bottom panel plot are color coded by SMBH mass.}
	\label{fig:thetalplots}	
\end{figure*}

Using our constraints on the coronal temperature, we constructed a compactness--temperature (\thetal) diagram using sources from our sample. $\theta$ is defined to be a dimensionless parameter for coronal temperature: $\theta = $ \kt/$m_{e}c^{2}$. In constructing the \thetal plane, we use \kt values obtained from the \xillver spectral model fit to the X-ray data. This eliminates the uncertainty in determining $\theta$ from \ecut values, since the conversion from \ecut to \kt varies depending on the optical depth of the corona \citep{petrucci-2001}, which can vary from source to source. In determining values of $l$, we assume a conservative value of 10$R_{g}$ for the coronal radius \citep{fabian-2015}. 

Figure~\ref{fig:thetalplots} presents compactness-temperature diagrams for the AGN in our sample for which black hole mass estimates are available. We plot several pair production lines corresponding to different coronal geometries. Treating the corona as an isolated cloud, \citet{svensson-1984} calculated the pair production line to have an analytical form $l \sim 10\theta^{5/2} e^{1/\theta}$, which is shown in solid cyan in our \thetal plots. We also mark pair production lines for a slab and hemispherical corona located above a reflecting accretion disk computed from \citet{stern-1995} (black and purple lines respectively in Figure~\ref{fig:thetalplots}). In the top panel of Figure~\ref{fig:thetalplots} the dashed lines correspond to the boundaries where electron-electron (black) and electron-proton (green) processes dominate over Compton cooling \citep{corona-heating-paper, fabian-1994}. The lower plot shows the distribution of sources on the \thetal plane individually color coded by black hole mass.     

These \thetal measurements for our large sample of \nustar-observed AGN generally show sources to be widely distributed in both temperature and radiative compactness, in contrast to previous measurements for smaller samples of AGN where sources appeared to cluster near the pair production lines \citep[e.g.,][]{fabian-2015,nikita-2018}. The results from \citet{ricci-2018}, which mapped the \thetal plane for 211 unobscured \bat AGN, show a distribution of individual sources fairly similar to our results. 


In general, the existence of AGN with low coronal temperatures is rather enigmatic, as the mechanisms behind coronal cooling in such sources are not well understood. Identification of AGN with low coronal temperatures is not uncommon, with findings of high energy cutoffs within the \nustar band reported in the literature in recent years \citep[e.g.,][]{kara-2017, yanjun-2017, nikita-2019}. Various theories have been proposed for possible cooling mechanisms to account for such low temperatures within the corona. Weak coronal heating mechanisms are a possibility, considering that it has been a longstanding open problem of supplying energy to the corona when it is established that the cooling timescale is shorter than the light crossing timescale \citep[e.g.,][]{corona-heating-paper,merloni-2001}. Low coronal temperatures could also be produced from a high optical depth within the corona. For optical depths exceeding unity, multiple inverse Compton scatterings of photons originating from the accretion disk could lead to effective coronal cooling \citep{kara-2017}. However, we have thus far assumed the corona is homogeneous, fully thermal, and at a single temperature, whereas in reality the corona is a dynamic structure and can have a range of temperatures \citep{fabian-2015}. It is possible that the corona is a hybrid plasma, containing both thermal and non-thermal particles \citep[e.g.,][]{Zdziarski-1993-comptonisation,corona-heating-paper}. \citet{fabian-2017} have shown through hybrid plasma simulations that a small fraction of non-thermal electrons with energies above 1 MeV can reduce the temperature of the corona, as electron-positron pairs redistribute their energy and reduce the mean energy per particle. 

Our \thetal measurements improve over previously reported studies by combining both a large sample size of AGN with high quality \nustar broadband X-ray data with improved spectral modeling for determination of coronal temperatures. By using the \xillvercp spectral model, we obtain more accurate estimates of \kt, since the high energy turnover in the intrinsic continuum is modeled as a Comptonized spectrum instead of the common exponential power law cutoff. Past works such as \citet{Zdziarski-1993} and \citet{fabian-2015} have shown that a simple exponential cutoff approximation produces a slower break in the X-ray spectrum compared to a Comptonized continuum, which retains a power law shape to higher X-ray energies before more rapidly turning over. Thus, an exponential cutoff approximation can lead to overestimates of the coronal temperature. 

We note that higher order effects can affect the precise location of sources on the \thetal plane. Most notably, general relativistic effects such as gravitational redshift and light bending can affect estimates of $l$ and $\theta$. For example, light bending can boost intrinsic values of $l$ and enhance Compton cooling, thereby moving pair lines to lower $\theta$. However, such corrections are highly dependent on properties such as disk inclination and coronal geometry, both of which are highly uncertain. Hence, we do not attempt to model general relativistic effects in this work due to the large uncertainties associated with such corrections.

\subsection{Optical Depth of the Coronal Plasma}

In this final section we report on the investigation of other AGN parameters derived from broadband X-ray spectral fit results for our full sample. Specifically, we focus on determining the optical depth of the plasma in the corona ($\tau$) and its relation to the Eddington ratio. The optical depth is not a parameter that is directly determined from X-ray spectral fitting, but it can be derived from $\Gamma$ and \kt according to the following equation given in \citet{nthcomp} for a spherical corona and formally valid for $\tau \gtrsim 1$:

\begin{equation}
\centering
\Gamma =  \sqrt{\frac{9}{4}+\frac{511~\mbox{keV}}{\tau kT_{e}(1+\tau/3)}} - \frac{1}{2}.
\label{eq:tau-eq}
\vspace{8pt}
\end{equation}

\begin{figure}[t]
	\hspace{-35pt}
	\includegraphics[width=0.56\textwidth]{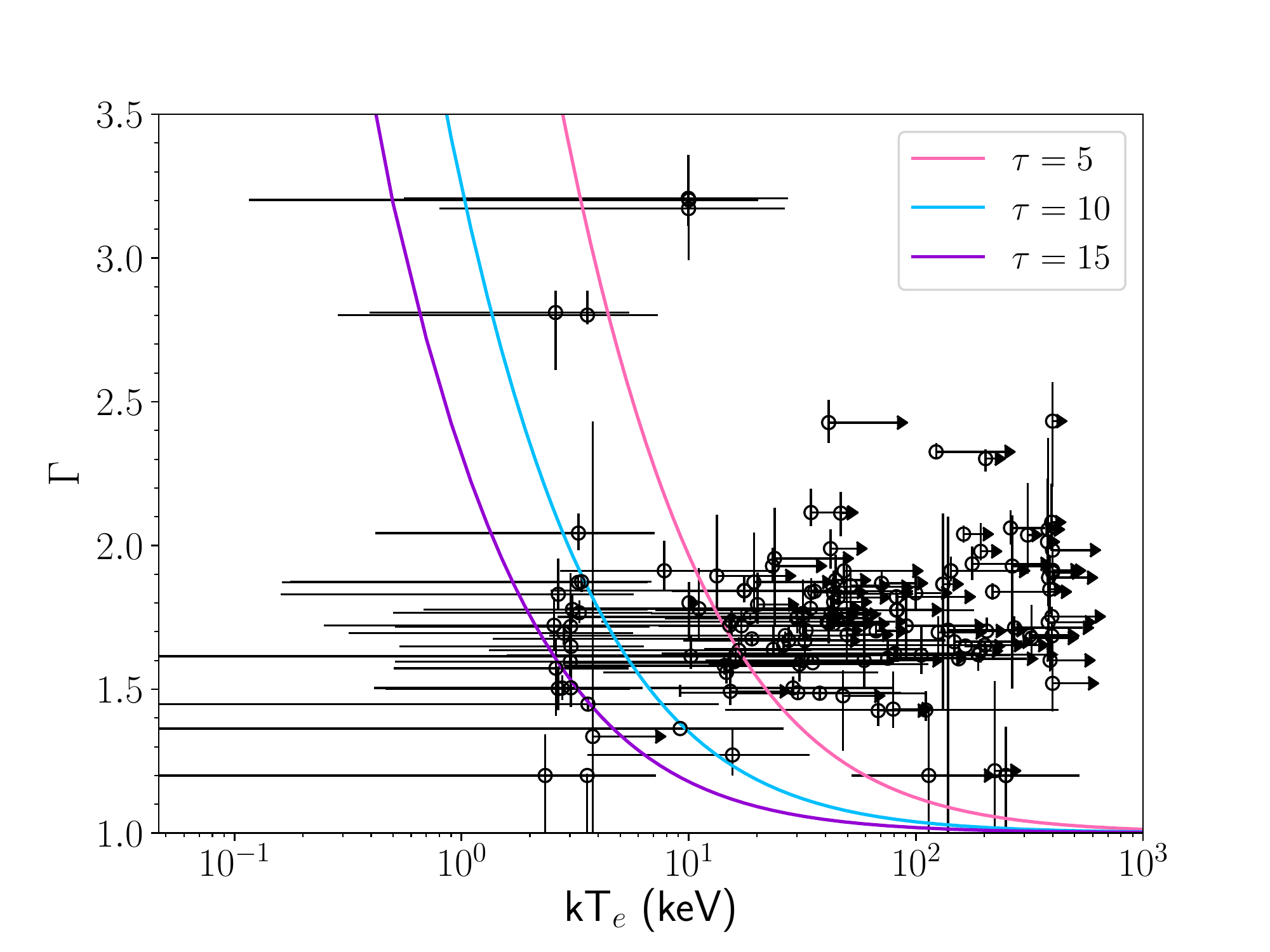}
	\caption{\kt and $\Gamma$ values obtained from the \xillver spectral model fit to the data. Solid colored lines correspond to curves of constant coronal plasma optical depth $\tau$, according to Equation~\ref{eq:tau-eq}.}
	\label{fig:gamma-kt-tau}	
\end{figure}

We use best-fit values of \kt found from the \xillvercp reflection model to solve for $\tau$ according to Equation~\ref{eq:tau-eq}. We find that the median and mean optical depths for our entire sample are $\tau =$ $3.04\pm1.73$ and $\tau =$ $4.84\pm1.80$ respectively. We note that a large number of sources in our sample have only a lower limit on \kt and so the values of $\tau$ presented in this work should be viewed as upper limits. We also note that parameter degeneracies, particularly in the \kt--$\Gamma$ plane, can lead to very high values of $\tau$. Spectral fits that produce very low values of \kt accompanied by low values of $\Gamma$ do not correspond to physically realistic conditions within the coronal plasma \citep[e.g.,][]{stern-1995,poutanen-1996}. In Figure~\ref{fig:gamma-kt-tau} we show best-fit values of \kt and $\Gamma$ along with curves of constant $\tau$ defined using equation~\ref{eq:tau-eq}. We observe that while there is some degree of degeneracy present in the \kt--$\Gamma$ plane, the majority of sources lie above the line roughly corresponding to the mean value of $\tau$ for our sample.   

For sources with full constraints on \kt, we performed Monte Carlo sampling as a rough estimate of the uncertainty in derived values of $\tau$. We drew 1000 random samples of \kt and $\Gamma$, assuming a mean and variance of the random sample taken from the observed respective distributions. We found no difference in the distributions of $\tau$ determined from the randomly sampled values of \kt and $\Gamma$ when assuming different underlying types of distribution (e.g. Gaussian vs. Poissonian). 

\begin{figure}[h!]
	\hspace{-35pt}
	\includegraphics[width=0.58\textwidth]{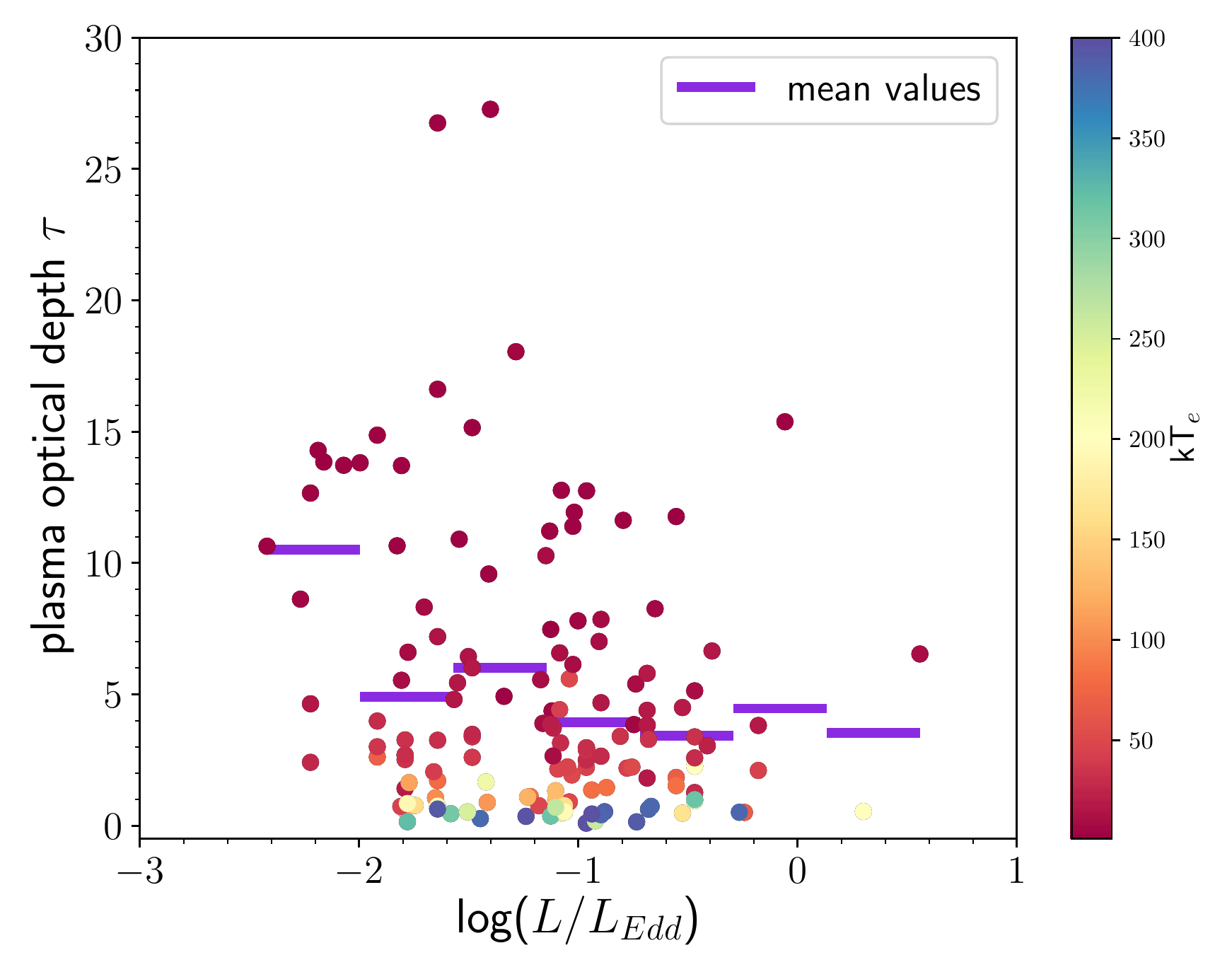}
	\caption{Coronal plasma optical depth $\tau$ versus the Eddington ratio \eddratio for our entire sample. Optical depth was derived from \kt and $\Gamma$ values obtained from the \xillver spectral model fit to the data.  Horizontal purple lines correspond to mean values of $\tau$ for different bins of Eddington ratio. Data points are color coded by their corresponding \kt value found from X-ray spectral fitting.}
	\label{fig:tauledd}	
\end{figure}

In Figure~\ref{fig:tauledd} we present results for $\tau$ against the Eddington ratio, with each source color-coded by its best-fit value of \kt. We do not find a statistically significant correlation between $\tau$ and \eddratio for individual data points or when binning the data by \eddratio, similar to results from \citet{ricci-2018}. From Figure~\ref{fig:tauledd} we also observe a trend of increasing optical depth with decreasing coronal temperature. \citet{tortosa-2018} also found a negative correlation between the plasma optical depth and coronal temperature for their sample of 19 \nustar-observed Seyfert 1 galaxies. This observed trend in $\tau$ with \eddratio supports the hypothesis mentioned in Section 4.4 for low temperature coronae possibly possessing high optical depths, thus enhancing coronal cooling.   

\pagebreak

\section{Summary}\label{sec:summary}

In this work, we have compiled a large sample of Seyfert 1 AGN with high quality broadband X-ray spectra taken with the \nustar observatory and studied fundamental properties of the coronal plasma that powers the continuum X-ray emission in AGN. We performed detailed broadband X-ray spectral modeling for all sources in our sample from which we obtained constraints on the temperature of the corona. From fitting a more physically accurate advanced reflection model, we find the mean coronal temperature to be \kt $=$ 84$\pm$9 keV, which is generally consistent with other measurements of high energy cutoffs for unobscured AGN reported in the literature. 

When investigating the relationship between the coronal temperature and accretion parameters such as the Eddington ratio and AGN SMBH mass, we do not find any strong correlations. We also examined the well-known \gammaledd relation, and found no statistically significant correlation, with little variation in the slope of the relation with the choice of X-ray spectral model used to determine $\Gamma$. We thus caution on the use of such relations previously presented in the literature to derive distributions of \eddratio or \mbh. 

We studied the distribution of sources in our sample across the compactness-temperature plane and find AGN to span a wide range of coronal temperatures and are not strictly confined to the boundary lines corresponding to runaway pair production. A number of sources appear to have fairly low coronal temperatures, which may arise from large optical depths of the coronal plasma, as we observe the optical depth of sources in our sample to increase with decreasing values of the coronal temperature. Another possibility is that the corona is a hybrid plasma system, where the presence of a population of non-thermal electrons can act to reduce the temperature of the plasma. Future studies that can apply advanced hybrid plasma models to high quality broadband AGN X-ray spectral observations performed with \nustar or concept X-ray missions such as \emph{HEX-P} \citep{kristin-hexp}, may be able to robustly test the possibility of such a physical scenario producing a low temperature corona.

\begin{acknowledgments}

We have made use of data from the \nustar mission, a project led by the California Institute of Technology, managed by the Jet Propulsion Laboratory, and funded by the National Aeronautics and Space Administration. We thank the \nustar Operations, Software and Calibration teams for support with the execution and analysis of these observations. This research has made use of the following resources: \nustar Data Analysis Software (NuSTARDAS) jointly developed by the ASI Science Data Center (ASDC, Italy) and the California Institute of Technology (USA); online \emph{Swift} data processing services provided by the SSDC; the High Energy Astrophysics Science Archive Research Center Online Service, provided by the NASA/Goddard Space Flight Center; NASA's Astrophysics Data System; {\tt Astropy}, a community-developed core Python package for Astronomy \citep{astropy-2013,astropy-2018}. Some of the optical spectra used for determining black hole masses used in this work were taken with the Doublespec (DBSP) instrument at Palomar via Yale (PI M. Powell, 2017-2019, 16 nights) as well as Caltech (PI F. Harrison) and JPL (PI D. Stern) from programs from 2013-2020.

M.\,Balokovi\'c acknowledges support from the Black Hole Initiative at Harvard University, which is funded in part by the Gordon and Betty Moore Foundation (grant GBMF8273) and in part by the John Templeton Foundation, as well as support from the YCAA Prize Postdoctoral Fellowship. M.K. acknowledges support from NASA through ADAP award NNH16CT03C. K.O. acknowledges support from the National Research Foundation of Korea (NRF-2020R1C1C1005462).
C.R. acknowledges support from the CONICYT+PAI Convocatoria Nacional subvencion a instalacion en la academia convocatoria a\~{n}o 2017 PAI77170080.

\end{acknowledgments}

\facilities{\nustar, \emph{Swift}, \xmm, \emph{Palomar Hale (DBSP)}}

\software{\tbabs \citep{tbabs}, \pexrav \citep{pexrav}, \nthcomp \citep{nthcomp}, \xillvercp \citep{xillver}, NuSTARDAS (v2.17.1), HEASOFT (v6.24), XMM SAS (v16.1.0), XSPEC (v12.8.2), Astropy \citep{astropy-2013,astropy-2018}, NumPy, Matplotlib \citep{matplotlib-2007}}

\vspace{15pt}

\bibliographystyle{aasjournal}
\bibliography{BAT_Sy1_refs}

\end{document}